%% file: MRea.tex
\documentclass[11pt,a4paper]{article}
\pdfoutput=1
\usepackage{jcappub}
\hypersetup{
    pdfauthor = {Rafael Marcondes},
    pdftitle = {Analytic study of the effect of dark energy-dark matter interaction on the growth of structures}
}
\usepackage{upgreek}
\usepackage{siunitx}
\usepackage{subfig}
\usepackage{booktabs}
\usepackage{multirow}
\usepackage[acronym,nomain,nonumberlist]{glossaries}
\input{glossary}

\glsdisablehyper

\title{Analytic study of the effect of dark energy-dark matter interaction on the growth of structures}
\author[a]{Rafael J.~F.~Marcondes,}
\author[a]{Ricardo C.~G.~Landim,}
\author[a]{Andr\'e A.~Costa,}
\author[b]{Bin Wang}
\author[a]{and Elcio Abdalla}

\affiliation[a]{Departamento de F\'isica Matem\'atica, Instituto de F\'isica,
    Universidade de S\~ao Paulo,\\
    Rua do Mat\~ao 1371, S\~ao Paulo, Brazil}
\affiliation[b]{Department of Physics and Astronomy, Shanghai Jiao Tong University \\
    200240 Shanghai, China}

\emailAdd{rafaelmarcondes@usp.br}
\emailAdd{rlandim@if.usp.br}
\emailAdd{alencar@if.usp.br}
\emailAdd{wang\_b@sjtu.edu.cn}
\emailAdd{eabdalla@if.usp.br}

\abstract{Large-scale structure has been shown as a promising cosmic probe for distinguishing and constraining dark energy models.
    Using the growth index parametrization, we obtain an analytic formula for the growth rate of structures in a coupled dark energy model in which the exchange of energy-momentum is proportional to the dark energy density.
    We find that the evolution of $f \sigma_8$ can be determined analytically once we know the coupling, the dark energy equation of state, the present value of the dark energy density parameter and the current mean amplitude of dark matter fluctuations.
    After correcting the growth function for the correspondence with the velocity field through the continuity equation in the interacting model,
    we use our analytic result to compare the model's predictions with large-scale structure observations.}
\keywords{dark energy theory, cosmological parameters from LSS}

\arxivnumber{1605.05264}

\DeclareSIUnit\parsec{pc}
\newcommand{\ud}{\text{d}}
\newcommand{\Nabla}{\nabla}
\newcommand{\de}{\text{\tiny DE}}
\newcommand{\dm}{\text{\tiny DM}}
\newcommand{\g}{\text{\tiny G}}
\newcommand{\m}{\text{\tiny M}}
\newcommand{\ba}{\text{\tiny B}}
\newcommand{\tot}{\text{\scriptsize total}}
\newcommand{\anl}{\text{\scriptsize anl}}
\newcommand{\tnum}{\text{\scriptsize num}}
\newcommand{\s}{\text{\scriptsize s}}

\begin{document}
    \setcounter{tocdepth}{2}
    \maketitle
    \flushbottom 

    \section{Introduction}
    Observations of \gls{sneIa} \cite{Riess1998,perlmutter1999} at the end of the 1990s culminated in the currently well established fact that the Universe is undergoing a phase of accelerated expansion.
    In the framework of \gls{gr}, some exotic form of matter with negative pressure --- the \gls{de} --- or a simple positive cosmological constant ($\Lambda$) can account for such acceleration.

    Prior to the idea of \gls{de} is another unknown component referred to as \gls{dm}.
    At the level of galaxy structures, the amount of visible matter in galaxies is not large enough to explain the observed rotation curves, which differ from the prediction of classical mechanics when considering the gravitational field generated by the visible matter \cite{Zwicky1933,Zwicky1937}. A possible solution is the existence of a kind of matter that neither interacts with radiation nor with the conventional matter except through the gravitational field or through some feeble interaction.

    The Universe turns out to be remarkably well described by a model composed mostly of these two dark components and smaller fractions of baryonic matter and radiation, the \gls{lcdm} model.
    This concordance \gls{lcdm} model, with a phase of extremely fast expansion (dubbed inflation) right after the Big Bang, is widely accepted as the de facto model of the Universe.
    But still, it leaves some questions unanswered.
    One of them is the seeming coincidence that matter and dark energy are found to contribute to the energy content of the Universe with amounts of the same order today, despite 
    behaving quite differently with respect to the expansion. 
    Posed this way, this fact has been known as the cosmic coincidence problem.
    Some authors have proposed the existence of a mechanism that drives the ratio between the two components close to 1 at late times.
    An interaction between the dark components would provide just that \cite{Amendola2000,delCampo2006,Poitras2014}, while appearing to be compatible with observations. 
    The possibility of such an interaction should be regarded as the natural case for these components, whose natures we do not know, rather than as a particular case of a bigger scenario.
    It could then be a solution or at least an alleviation to the coincidence problem.

    Several authors have built and constrained interacting \gls{de} models with observations, mostly of \gls{sneIa}, \gls{cmb}, \gls{bao}, galaxy clusters and $H(z)$ data \cite{HeWang2008,Abdalla2009,Abdalla2010,HeWangAbdalla2011,CaoLiangZhu2011,SolanoNucamendi2012,Li2013,CaoLiang2013,AndreElcio2014PRD89,Andre2014,LeDelliouetal2015}.
    For more comprehensive references, see the review ref.~\cite{Review2016}.
    More recently, researchers have been also attempting to detect the effect of interaction on the clustering of matter, through the rate at which structures form, which can be measured with \glspl{rsd} \cite{Tsujikawa2013,YangXu2014apr,Li2014,Richarte2015,Li2015}.
    The fact that the interaction is expected to affect the growth of structures more than it affects phenomena from remote epochs, e.g.~the \gls{cmb}, makes this low-redshift observable particularly interesting.
    One can expect to see the imprint of the interaction on matter structures by analyzing the rate at which they grow compared to how the non-interacting standard cosmology predicts.

    It is interesting that the evolution of the growth rate can be solved approximately in an analytic form, $f(z) \approx \left[ \Omega_{\m}(z) \right]^{\gamma(z)}$.
    The approximation was first proposed by Peebles \cite{Peebles1980} for the matter dominated universe as
    $f(z=0) \approx \left( \Omega_{\m,0} \right)^{0.6}$, followed by the more accurate approximation $\left( \Omega_{\m,0} \right)^{4/7}$ by Lightman \& Schechter \cite{Lightman1990}.
    More generally,
    the approximation was also obtained in dynamical \gls{de} models with zero curvature and slowly varying \glsentrylong{eos} \cite{WangSteinhardt1998}
    and in curved spaces \cite{Gong2009}.
    In modified gravity models, the approximate solution was given in refs.~\cite{Linder2007,Gong2008}.
    Since growth data spans a wide range of redshift and the growth index evolves with the redshift, it is worth exploring its parametrization as a function of the redshift. This can help distinguish between \gls{de} models and modified gravity models \cite{GannoujiPolarski2008,Dossett2010}.

    In this work we will investigate the influence of a \gls{de}-\gls{dm} interaction on the growth of structures.
    Our main purpose is to solve analytically the growth rate of matter perturbations as a function of the redshift in an interacting \gls{de} model.
    We will generalize the method employed for the dynamical \gls{de} model without any interaction with \gls{dm} in ref.~\cite{Tsujikawa2013}.
    Our derivation is based on the expansion of the growth index and of the \gls{de} \glsentryfull{eos} parameter in terms of the \gls{de} density parameter $\Omega_{\de}(z)$.
    We will also derive an expression for the \glsentryfull{rms} amplitude of perturbations $\sigma_8(z)$ and show that when
    the \gls{de} equation of state, the coupling, the \gls{de} energy density and the amplitude of perturbations at present are given, the evolution history of the growth of structures is fully determined analytically.
    This analytic solution of the growth can help us clearly see the influence of the interaction between dark sectors in the growth.
    With the analytic form of $f\sigma_8(z)$ obtained, we can test the interacting \gls{de} model by using \gls{rsd} observations.

    The outline of this paper is as follows: in section~\ref{sec:models} we introduce the phenomenological model.
    Section~\ref{sec:analysis} goes deeply into the dynamical equations governing the perturbations of the fluids, according to \gls{gr}. We then compare our model predictions with \glspl{rsd} in section~\ref{sec:data} and give our conclusions in section~\ref{sec:conclusions}.

    \section{Cosmology with \glsentrytext{de} and \glsentrytext{dm} interactions}
    Although some models inspired by Quantum Field Theory have been proposed \cite{FarrarPeebles2004,Andre2014} attempting to explain the interaction between dark sectors at the Lagrangian level, here we will concentrate on the phenomenological approach to describe the interaction between \gls{de} and \gls{dm}. We represent the interaction by non-vanishing contributions to the right-hand side of the energy-momentum tensor conservation equations for the dark fluids, preserving, however, the total energy-momentum conservation.
    \label{sec:models}

    \subsection{The background universe}
    Hereafter we consider a universe described, at the background level, by the flat \gls{flrw} metric, which we write in terms of the conformal time $\tau$, $\ud s^2 = a^2(\tau) \left( - \ud \tau^2 + \delta_{ij} \, \ud x^i \, \ud x^j \right)$, where $a(\tau)$ is the scale factor and $x^i$ are the spatial coordinates.
    We use latin letters $i, j, \ldots$ for the spatial indices $1, 2, 3$ and greek letters $\mu, \nu, \ldots$ for the indices \numlist{0;1;2;3}. Dots denote derivatives with respect to the conformal time.

    We consider the universe composed only of dark matter and dark energy, with zero curvature. The components are treated as fluids, with energy-momentum tensor
    \begin{align}
        \label{eq:EMtensor}
        \bar T_{\mu\nu} = \bar p \bar g_{\mu\nu} + (\bar p + \bar \rho) \bar u_{\mu} \bar u_{\nu},
    \end{align}
    where $\bar p$ and $\bar \rho$ are the pressure and energy density of the fluid, $\bar u^{\mu}$ its four-velocity, and $\bar g_{\mu\nu}$ the metric. 
    The bars indicate that the quantities are unperturbed.
    In the standard model, the fluids satisfy the energy-momentum conservation $\Nabla_{\mu} {\bar{T}^{\mu}}_{\phantom{\mu}\nu} = 0$. 
    An interaction is introduced by rewriting this equation as
    \begin{subequations}
        \begin{align}
            \label{eq:dmequation}
            \Nabla_{\mu} {\bar T}^{\mu\phantom{\nu}\dm}_{\phantom{\mu}\nu} &= {\bar Q}^{\phantom{\nu}\dm}_{\nu}, \\
            \Nabla_{\mu} {{\bar{T}^{\mu\phantom{\nu}\de}}_{\phantom{\mu}\nu}} &= {\bar Q}^{\phantom{\nu}\de}_{\nu},
        \end{align}
    \end{subequations}
    for both fluids. Total energy-momentum conservation requires ${\bar{Q}_{\nu}}^{\phantom{\nu}\de} = -{\bar{Q}_{\nu}}^{\phantom{\nu}\dm}$. 
    As a result of the homogeneity and isotropy of the background, the spatial components of $\bar{Q}_{\nu}$ are zero.
    The fluids are comoving with the Hubble flow, with $\bar u^{\mu} = (a^{-1}, 0, 0, 0)$.
    The unperturbed energy-momentum tensors of the two fluids have their non-zero components given by ${\bar{T}^{0}_{\phantom{0}0}} = - \bar\rho$, ${\bar T^i_{\phantom{i}j}} = \bar p \, \delta^i_j$.
    The $\nu=0$ energy-momentum conservation equation thus reads
    \begin{align}
        \label{eq:Bianchi0background}
        \dot{\bar\rho} + 3 \mathcal{H} \left(1 + w\right) \bar\rho = a^2 \bar Q^0 = - \bar Q_0,
    \end{align}
    for each of the two dark fluids, with $\mathcal{H} \equiv \dot a/a$ and $w \equiv \bar p/\bar \rho$ the \gls{eos} parameter.
    The background evolution of the universe as a whole is governed by the Friedmann equation, from the unperturbed time-time \gls{efe},
    \begin{align}
        \label{eq:Friedmann}
        \mathcal{H}^2 &= \frac{8 \uppi G}{3} a^2 \left( \bar \rho_{\dm} + \bar \rho_{\de} \right).
    \end{align}

    \subsection{The perturbed equations}
    \label{ss:pertequations}
    We consider scalar perturbations only. 
    Since we intend to discuss the gravitational evolution of perturbations, it is more convenient to work in the conformal (Newtonian) gauge. 
    Scalar perturbations develop at low redshifts in the era of structure formation.
    In the conformal gauge, the line element is written as
    \begin{align}
        \ud s^2 &= a^2(\tau) \left[ -\left(1+2\psi\right) \ud\tau^2 + \left(1 - 2\phi\right) \delta_{ij} \ud x^i \ud x^j \right],
    \end{align}
    with $\phi = \phi(x^{\mu})$ and $\psi = \psi(x^{\mu})$ being small perturbations, satisfying $|\phi|, |\psi| \ll 1$.
    The metric is given by
    \begin{align}
        g_{00} = - a^2\left(1 + 2\psi\right), \qquad
        g_{0i} = g_{i0} = 0, \qquad
        g_{ij} = a^2 \left(1 - 2 \phi \right) \delta_{ij}.
    \end{align}
    Assuming that there is no anisotropic stress, we have $\psi = \phi$.
    We rewrite the metric separating its unperturbed part and the perturbation $h_{\mu\nu}$ as $g_{\mu\nu} = \bar g_{\mu\nu} + h_{\mu\nu}$.
    We have then $\bar g_{00} = -a^2$, $\bar g_{ij} = a^2 \delta_{ij}$, $h_{00} = -2 a^2 \phi$, $h_{ij} = -2 a^2 \phi \delta_{ij}$. The unperturbed metric $\bar g$ must be used to lower or raise indices of unperturbed tensors.
    We denote the perturbed parts of all other quantities by their own symbol preceded by a $\delta$, as in $B \equiv \bar B + \delta B$.
    The perturbations must satisfy $|\delta B| \ll |\bar B|$.

    The components of the perturbed energy-momentum tensors, from perturbing eq.~\eqref{eq:EMtensor}, are
    \begin{gather}
        \begin{gathered}
            \delta {T^0}_0 = -\delta \rho, \qquad \qquad
            \delta {T^i}_j = \delta p \, \delta^i_j, \qquad \qquad
            \delta T_{00} = a^2 \left( \delta \rho + 2 \bar\rho \phi \right), \\
            \delta {T^i}_0 = - a^{-1} \left(\bar \rho + \bar p \right) \delta u^i, \qquad \qquad
            \delta {T^0}_i = a^{-1} \left(\bar \rho + \bar p \right) \delta u_i, \\
            \delta T_{0i} = \delta T_{i0} = - a \left( \bar \rho + \bar p \right) \delta u_i, \qquad \qquad
            \delta T_{ij} = a^2 \left( \delta p - 2 \bar p \phi \right) \delta_{ij}. 
        \end{gathered}    
    \end{gather}
    With these perturbed metric and energy-momentum tensor, the perturbed conservation equations give the following evolution equations for the perturbations,
    \begin{subequations}
        \label{eq:dynamicalequations}
        \begin{align}
            - \dot\delta - \left[3 \mathcal{H} \left( c_s^2 - w \right) - \frac{\bar Q_{0}}{\bar\rho} \right]  \delta  - \left(1 +w \right) \bigl(\theta - 3 \dot \phi\bigr) &= \frac{{\delta Q}_{0}}{\bar\rho}, \\
            \dot\theta + \left[\mathcal{H} \left(1 -3w \right) - \frac{\bar Q_{0}}{\bar\rho} + \frac{\dot w}{1+w} \right] \theta - k^2 \phi - \frac{c_s^2}{1+w} k^2 \delta &= \frac{i k^i{\delta Q}_{i}}{\bar\rho \left(1+w\right)},
        \end{align}
    \end{subequations}
    where we have introduced the relative density perturbation $\delta \equiv \delta \rho/\bar \rho$ and used $\delta u_0 = -a \phi$, from the condition $g_{\mu\nu} u^{\mu} u^{\nu} = -1$.
    $\delta Q_{\mu}$ are the perturbations to the exchange of energy-momentum in the perturbed conservation equations,
    $c_s^2 \equiv \delta p/\delta \rho$ is the sound speed of the fluid, $k^i$ are the components of the wavevector in Fourier space, and $\theta \equiv a^{-1} i k^j \delta u_j$ is the divergence of the velocity perturbation in Fourier space. 

    The perturbed time-time \gls{efe} is the Poisson equation, relating $\phi$ and the total density perturbation of the fluids. 
    The full Poisson equation in Fourier space is
    \begin{align}
        \label{eq:PoissonFullEq}
        \left(1 + 3\mathcal{H}^2/k^2 \right) k^2 \phi &= -3 \mathcal{H} \dot\phi -4 \uppi G a^2 \left( \bar\rho_{\dm} \delta_{\dm} + \bar\rho_{\de} \delta_{\de} \right).
    \end{align}
    In order to analyze the growth of structures, we need to combine equations \eqref{eq:dynamicalequations} with the Poisson equation to substitute $\phi$ in terms of $\delta_{\dm}$.
    Since structures grow in the Newtonian regime, on spatial scales much smaller than the horizon ($k \gg \mathcal{H}$) and with negligible time variation of the potential, we can discard the second term in the left-hand side of eq.~\eqref{eq:PoissonFullEq} and the term proportional to $\dot\phi$.
    Also, the dark energy perturbations are expected to be negligible on sub-horizon scales \cite{Maartens2013}.
    The Poisson equation then reduces to
    \begin{align}
        \label{eq:Poisson}
        k^2 \phi &= -4 \uppi G a^2 \bar\rho_{\dm} \delta_{\dm} 
        = - \frac{3}{2} \mathcal{H}^2 \Omega_{\dm} \delta_{\dm},
    \end{align}
    the last equality coming from eq.~\eqref{eq:Friedmann} with the density parameter of the \gls{dm} fluid defined as $\Omega_{\dm} \equiv \bar\rho_{\dm}/\bar\rho_{\text{\scriptsize critical}}$, the critical density being equal to the total density $\bar\rho_{\tot} = \bar\rho_{\dm} + \bar\rho_{\de}$ (and thus $\Omega_{\de} = 1 - \Omega_{\dm}$) in the absence of curvature.
    Finally, combining equations \eqref{eq:dynamicalequations} and \eqref{eq:Poisson} together, we get the second order differential equation for the \gls{dm} perturbation
    \begin{align}
        \label{eq:fullSOE}
        \ddot \delta_{\dm} &- \left( \mathcal{Q} - \mathcal{K} \right) \dot\delta_{\dm} - \left( \frac{3}{2} \mathcal{H}^2 \Omega_{\dm} + \dot{\mathcal{Q}} + \mathcal{K} \mathcal{Q} \right) \delta_{\dm} =
         - \frac{i k^i \delta Q_i^{\phantom{i}\dm}}{\bar\rho_{\dm}},
    \end{align}
    with
    \begin{align}
        \mathcal{Q} \equiv \frac{{\bar Q_0}^{\phantom{0}\dm}}{\bar\rho_{\dm}} - \frac{\delta {Q_0}^{\dm}}{\bar\rho_{\dm} \delta_{\dm}}
        \qquad \text{and} \qquad \mathcal{K} \equiv \mathcal{H} - \frac{{\bar Q_0}^{\phantom{0}\dm}}{\bar\rho_{\dm}}.
    \end{align}
    Eq.~\eqref{eq:fullSOE} is general and valid for any type of interaction.
    In the next subsection we simplify this equation by choosing a particular coupled model.

    \subsection{The phenomenological coupled \glsentrytext{de} model and the DM evolution}
    \label{ss:intromodel2}
    We analyze a model with an interaction term in the \gls{dm} energy-momentum conservation equation that is proportional to the \gls{de} energy density, 
    \begin{align}
        \text{\glsentryshort{cde}:} \qquad  \qquad {Q_0}^{\dm} &= {\bar
            Q}_0^{\phantom{0}\dm} = -3 \mathcal{H} \xi \bar\rho_{\de}.
        \label{eq:model2Q0}
    \end{align}
    Models like this, with interaction proportional to $\rho_{\de}$, $\rho_{\dm}$ or their combination, have been extensively studied in the last years, for example in refs.~\cite{Abdalla2009,Abdalla2010,HeWangAbdalla2011,CaoLiangZhu2011,CaoLiang2013,AndreElcio2014PRD89} (see also the recent review \cite{Review2016}).
    The interaction in this \gls{cde} model has only an unperturbed part, since we are neglecting \gls{de} clustering.
    With eq.~\eqref{eq:model2Q0}, the background evolution eq.~\eqref{eq:Bianchi0background} reads
    \begin{align}
        \label{eq:BianchiDMmodel2}
        \dot {\bar{\rho}}_{\dm} + 3 \mathcal{H} \bar\rho_{\dm} = 3 \mathcal{H} \xi \bar\rho_{\dm} \tfrac{1 - \Omega_{\dm}}{\Omega_{\dm}}.
    \end{align}
    Replacing $\frac{\bar Q_0^{\phantom{0}\dm}}{\bar\rho_{\dm}} = - 3 \mathcal{H} \xi \frac{1 - \Omega_{\dm}}{\Omega_{\dm}}$ and $\frac{\delta Q_0^{\phantom{0}\dm}}{\bar\rho_{\dm}\delta_{\dm}} = 0$, 
    eqs.~\eqref{eq:dynamicalequations} for \gls{dm} are
    \begin{subequations}
        \begin{align}
            \label{eq:DMcont}
            \dot \delta_{\dm} + 3 \mathcal{H} \xi \tfrac{1 - \Omega_{\dm}}{\Omega_{\dm}}  \delta_{\dm} + \theta_{\dm} &= 0, \\
            \dot \theta_{\dm} + \mathcal{H} \left( 1 + 3 \xi \tfrac{1 - \Omega_{\dm}}{\Omega_{\dm}} \right) \theta_{\dm} +  \tfrac{3}{2} \mathcal{H}^2 \Omega_{\dm} \delta_{\dm} &= 0,
        \end{align}
    \end{subequations}
    and the evolution of the \gls{dm} perturbations \eqref{eq:fullSOE} reduces to
    \begin{align}
        \ddot \delta_{\dm} 
        &+ \left(1 + 6 \xi \tfrac{1-\Omega_{\dm}}{\Omega_{\dm}} \right) \mathcal{H} \dot \delta_{\dm} - {} \nonumber \\
       \label{eq:Model2eq}
       &-\tfrac{3}{2} \mathcal{H}^2 \delta_{\dm} \left[
           \Omega_{\dm}
           - 2 \xi \tfrac{1 - \Omega_{\dm}}{\Omega_{\dm}} \left(
               1
               + \tfrac{\dot{\mathcal{H}} }{\mathcal{H}^2}
               + 3 \xi \tfrac{1-\Omega_{\dm}}{\Omega_{\dm}} 
               - \tfrac{\dot\Omega_{\dm}}{\mathcal{H}\Omega_{\dm}} \tfrac{1}{1 - \Omega_{\dm}} 
   \right) \right]
       = 0.
    \end{align}
    The standard evolution $\ddot \delta_{\dm} + \mathcal{H} \dot \delta_{\dm} - \tfrac{3}{2} \mathcal{H}^2 \Omega_{\dm} \delta_{\dm} = 0$ is recovered when $\xi = 0$.
    Due to the presence of the interaction, the coefficient of $\delta_{\dm}$ in eq.~\eqref{eq:Model2eq} can become positive as $\Omega_{\dm}$ decreases, leading to a decaying regime of the perturbation.
    This negative growth rate, as we will see in section~\ref{sec:analysis}, cannot be described by the parametrization of $f$ with the growth index.

    \section{The analytical growth rate and amplitude of perturbations}
    \label{sec:analysis}
    The growth rate $f$ is defined as the logarithmic derivative of the (total) matter perturbation with respect to the logarithm of the scale factor,
    \begin{align}
        f \equiv \frac{\ud \ln \delta_{\m}}{\ud \ln a},
    \end{align}
    and is often parametrized by 
    \begin{align}
        \label{eq:fOmegagamma}
        f \approx \Omega_{\m}^{\gamma}.
    \end{align}
    The exponent $\gamma$ is called the growth index.
    This approximation has been shown very satisfactory until now for virtually any cosmological model without \gls{de}-\gls{dm} coupling, with $\gamma$ varying accordingly (see, for example, ref.~\cite{Linder2005} and references therein).
    In the $\Lambda$CDM model, the growth index is approximately \num{6/11}.
    
    One can explore $\gamma$ in different models by expanding it in terms of the \gls{de} density parameter.
    This can then be used to test and compare models since we can measure the growth rate as a function of the redshift.

    \subsection{The growth of structure in the \glsentrytext{cde} model}
    \label{ss:CDEmodel}
    To obtain the approximation $f \approx \Omega_{\dm}^{\gamma}$, we need to change the time derivatives $\partial/\partial \tau$ to $\partial/\partial a$ and write eq.~\eqref{eq:Model2eq} in terms of $f$.
    We can carry out the Taylor expansion for the functions in terms of $\Omega_{\de}$ around zero, describing the time evolution in terms of the \gls{de} density abundance.
    In non-interacting models, a polynomial equation in $\Omega_{\de}$ can be obtained by equating coefficients in both sides, with its zero-th order coefficients vanishing identically and its coefficients for higher orders in $\Omega_{\de}$ giving the coefficients of $\gamma = \sum_{n=0}^{\infty} \gamma_n \left( \Omega_{\de}\right)^n$ in terms of the coefficients of $w_{\de} = \sum_{n=0}^{\infty} w_n \left( \Omega_{\de}\right)^n$ (see, for example, ref.~\cite{WangSteinhardt1998}).
    This form of parametrization has been shown useful in obtaining the analytic expression of the growth index in dynamical \gls{de} models and convenient for distinguishing the model from the $\Lambda$CDM model \cite{WangSteinhardt1998,Tsujikawa2013}. 

    For the \gls{de}-\gls{dm} interaction model, we will adopt the same strategy as that of the non-interacting cases \cite{Tsujikawa2013}.
    We will do the expansion around $\Omega_{\de} = 0$ and assume that the ratio between the rate of change of the \gls{de} density parameter and the Hubble rate is negligible compared to the density parameter and to unity, at least in the regime of structure formation.
    Therefore, $\dot\Omega_{\de}/\mathcal{H} \ll \Omega_{\de}$ in eq.~\eqref{eq:Model2eq} and we are led to
    \begin{align}
        \ddot \delta_{\dm} 
        &+ \left(1 + 6 \xi \tfrac{1-\Omega_{\dm}}{\Omega_{\dm}} \right) \mathcal{H} \dot \delta_{\dm} - {} \nonumber \\
        \label{eq:ddotdeltaModel2}
        &-\tfrac{3}{2} \mathcal{H}^2 \delta_{\dm} \left\lbrace
           \Omega_{\dm}
           + 2 \xi \tfrac{1 - \Omega_{\dm}}{\Omega_{\dm}} 
           \left[
               -\tfrac{1}{2} + 3 w_{\de} \left( 1 - \Omega_{\dm} \right) 
               - 3 \xi \tfrac{1 - \Omega_{\dm}}{\Omega_{\dm}}
   \right] 
   \right\rbrace
       = 0.
    \end{align}
    After some manipulations, this is rewritten as
    \begin{align}
        \frac{\ud^2 \ln \delta_{\dm}}{\ud \ln a^2} + \left( \frac{\ud \ln \delta_{\dm}}{\ud \ln a} \right)^2 &+ \left[ \frac{1}{2} - \frac{3}{2} w_{\de} \left(1 - \Omega_{\dm} \right) + 6 \xi \frac{1 - \Omega_{\dm}}{\Omega_{\dm}} \right] \frac{\ud \ln \delta_{\dm}}{\ud \ln a} - {} \nonumber \\
        {}& - \frac{3}{2} \Omega_{\dm} + 3 \xi \frac{1 - \Omega_{\dm}}{\Omega_{\dm}} \left[\frac{1 - \Omega_{\dm}}{\Omega_{\dm}} \left( 3 \xi - 3 w_{\de} \Omega_{\dm} \right) + \frac{1}{2} \right] = 0.
    \end{align}
    Substituting $f$, we have
    \begin{align}
        \label{eq:dfdlnaM2}
        \frac{\ud f}{\ud \ln a} + f^2 + f & \left[ \frac{1}{2} - \frac{3}{2} w_{\de} \left(1 - \Omega_{\dm} \right) + 6 \xi \frac{1-\Omega_{\dm}}{\Omega_{\dm}} \right] - \frac{3}{2} \Omega_{\dm} + {} \nonumber \\
        {}& + 3 \xi \frac{1 - \Omega_{\dm}}{\Omega_{\dm}} \left[\frac{1 - \Omega_{\dm}}{\Omega_{\dm}} \left( 3\xi - 3 w_{\de} \Omega_{\dm} \right) + \frac{1}{2} \right] = 0,
    \end{align}
    which still has the first term parametrized by the scale factor.
    Next, we write $\frac{\ud f}{\ud \ln a} = \frac{\ud \Omega_{\dm}}{\ud \ln a} \frac{\ud f}{\ud \Omega_{\dm}}$ and use the (background) energy conservation equations to substitute $\frac{\ud \Omega_{\dm}}{\ud \ln a}$.
    The total conservation equation gives
    \begin{gather}
        \ud \bar \rho_{\tot} + 3 \tfrac{\ud a}{a} \left(\bar \rho_{\tot} + \bar p_{\tot} \right) = 0, \nonumber \\
        \ud (a^3 \bar \rho_{\tot}) = - \ud (a^3) w_{\de} \bar\rho_{\de}, \nonumber \\
        \label{eq:totcons}
        \ud \left( \frac{a^3 \bar\rho_{\dm}}{\Omega_{\dm}} \right) = - \ud (a^3) w_{\de} \bar \rho_{\dm} \frac{1-\Omega_{\dm}}{\Omega_{\dm}},
    \end{gather}
    where we have used $\bar\rho_{\tot} = \frac{\bar\rho_{\dm}}{\Omega_{\dm}}$ and $\bar\rho_{\de} = \bar\rho_{\dm} \frac{1-\Omega_{\dm}}{\Omega_{\dm}}$, while the \gls{dm} equation gives
    \begin{gather}
        \ud \bar \rho_{\dm} + 3 \tfrac{\ud a}{a} \bar \rho_{\dm} = 3 \tfrac{\ud a}{a} \xi \bar\rho_{\dm} \tfrac{1 - \Omega_{\dm}}{\Omega_{\dm}}, \nonumber \\
        \ud(a^3 \bar\rho_{\dm}) = \xi \bar\rho_{\dm} \tfrac{1 - \Omega_{\dm}}{\Omega_{\dm}}  \ud(a^3),
    \end{gather}
    which can be inserted back in eq.~\eqref{eq:totcons} to give
    \begin{gather}
        \xi \bar \rho_{\dm} \frac{1 - \Omega_{\dm}}{\Omega_{\dm}} \frac{\ud (a^3)}{\Omega_{\dm}} - a^3 \bar\rho_{\dm} \frac{\ud \Omega_{\dm}}{\Omega_{\dm}^2} = - w_{\de} \bar \rho_{\dm} \frac{1 - \Omega_{\dm}}{\Omega_{\dm}} \ud (a^3), \nonumber \\
        3 \xi \left(1 - \Omega_{\dm} \right) \ud \ln a - \ud \Omega_{\dm} = - 3 w_{\de} \Omega_{\dm} \left(1 - \Omega_{\dm} \right) \ud \ln a, \nonumber \\
        \label{eq:dOdlnaM2}
        \frac{\ud \Omega_{\dm}}{\ud \ln a} = 3 \left(1 - \Omega_{\dm} \right) \left( \xi + w_{\de} \Omega_{\dm} \right).
    \end{gather}
    Substituting eq.~\eqref{eq:dOdlnaM2} into $\frac{\ud f}{\ud \ln a} = \frac{\ud \Omega_{\dm}}{\ud \ln a} \frac{\ud f}{\ud \Omega_{\dm}}$ and dividing eq.~\eqref{eq:dfdlnaM2} by $f$ we have
    \begin{align}
        3 \left( \xi + w_{\de} \Omega_{\dm} \right) \frac{1 - \Omega_{\dm}}{f} & \frac{\ud f}{\ud \Omega_{\dm}} + f + \frac{1}{2} - \frac{3}{2} w_{\de} \left(1 - \Omega_{\dm} \right) + 6 \xi \frac{1 - \Omega_{\dm}}{\Omega_{\dm}} - \frac{3}{2} \frac{\Omega_{\dm}}{f} + {} \nonumber \\
        \label{eq:fullfModel2}
        {} &+ 3 \xi \frac{1 - \Omega_{\dm}}{f \Omega_{\dm}} \left[ \frac{1 - \Omega_{\dm}}{\Omega_{\dm}} \left( 3 \xi - 3 w_{\de} \Omega_{\dm} \right) + \frac{1}{2} \right] = 0.
    \end{align}
    Finally, expanding eq.~\eqref{eq:fullfModel2} around $\Omega_{\de} = 0$ with $f = \left(\Omega_{\dm}\right)^{\gamma_0 + \gamma_1 \Omega_{\de} + \ldots}$, we arrive at the polynomial equation
    \begin{align}
        \label{eq:poleqCDE}
        \left[ 3 \left(1 - w_0 + 5 \xi \right) - \gamma_0 \left( 5 - 6 w_0 - 6 \xi \right) \right] \Omega_{\de}  
            + \frac{1}{2} \left[ 
            -\gamma_0^2
            + \gamma_0 \left(
                        1
                        +12 w_1
                        + 18 \xi 
                    \right)
                    - {} \right. \nonumber \\
                    {} - \left. \vphantom{\gamma_0^2}  2 \gamma_1 \left(
                5 
                - 12 w_0
                - 12 \xi
                    \right)
                - 6w_1
                + 6 \xi \left( 
                    5
                    - 6 w_0
                    +6 \xi
                \right)
            \right] \Omega_{\de}^2
        + \mathcal{O} (\Omega_{\de}^3) = 0.
    \end{align}
    The zero-th order part is still identically zero even with non-zero $\xi$.
    The equations of the higher order terms can be solved to give the modified growth index coefficients
    \begin{subequations}
        \label{eq:gammas}
        \begin{align}
            \gamma_0 &= \frac{3 \left(1 - w_0 + 5 \xi \right)}{5 - 6 w_0 - 6 \xi}, \\
            \gamma_1 &= \frac{- \gamma_0^2 + \gamma_0 \left(1 + 12 w_1 + 18 \xi \right) - 6 w_1 + 6 \xi \left(5 - 6 w_0 + 6 \xi \right)}{2 \left( 5 - 12 w_0 -12 \xi\right)}, \\
            \vdots \nonumber
        \end{align}
    \end{subequations}
    Eqs.~\eqref{eq:gammas} allow us to analyze the effect of the interaction and of the \gls{eos} on the growth index.
    We note that positive $\xi$ increases $\gamma_0$, the dominant part of the growth index.
    For example, a $\xi = \pm 0.01$ coupling changes $\gamma_0$ by approximately \SI{\pm 3}{\percent} and the measured growth by up to \SI{\pm 9}{\percent} at $z = 0$ when $w_0 = -1$ and $\sigma_{8,0}$ and $\Omega_{\de,0}$ are also fixed at fiducial $\Lambda$CDM values (see also figure~\ref{fig:differentxi}). 
    The well-known result $\gamma_0 = \frac{3 \left(1 - w_0 \right)}{5 - 6 w_0}$ is recovered when $\xi= 0$, giving $\gamma_0 = 6/11$ for $\Lambda$CDM.
    With the standard values $w_0 = -1$ and $\xi = 0$, the first-order coefficient is $\gamma_1 = \frac{3 \left(5 + 11 w_1 \right)}{2057}$, which gives a rather small contribution $\gamma_1 \Omega_{\de}$ to $\gamma$ for a slowly varying \gls{eos} parameter.

    Predictions made with $f = \left(\Omega_{\dm}\right)^{\gamma_0 + \gamma_1 \Omega_{\de} + \ldots}$ can, in principle, be compared to growth rate measurements like those compiled in ref.~\cite{Dossett2010}.
    Those data, however, are generally obtained from measurements of the \gls{rsd} parameter $\beta = f/b$, where $b$ is the bias measuring how galaxies trace the matter density field, and thus can be bias-dependent. 
    Usually, it is preferable to compare predictions with the bias-independent data of the combination $f \sigma_{8}$ \cite{SongPercival2009}, the growth rate multiplied by the variance of the density field filtered at a scale $R = 8\,h^{-1}\,\si{\mega\parsec}$, defined as
    \begin{align}
        \label{eq:s8def}
        \sigma_{R}^2 (z) \equiv \frac{1}{2 \uppi^2} \int_0^{\infty} \ud k \, k^2 P(k, z) \left| W( k R) \right|^2,
    \end{align}
    where $P(k, z)$ is the matter power spectrum and $W(kR)$ is the window function of the experiment in Fourier space.
    We derive $\sigma_8$ from $\delta_{\dm}$ starting with the definition of $f$,
    \begin{gather}
        \frac{\ud \Omega_{\dm}}{\ud \ln a} \frac{\ud \ln \delta_{\dm}}{\ud \Omega_{\dm}} = (\Omega_{\dm})^{\gamma}  \quad \Rightarrow \quad 3 \left(1 - \Omega_{\dm} \right) \left(\xi + w_{\de} \Omega_{\dm} \right) \frac{\ud \ln \delta_{\dm}}{\ud \Omega_{\dm}} = (\Omega_{\dm})^{\gamma} \quad \therefore \nonumber \\
        \therefore \quad \frac{\ud \ln \delta_{\dm}}{\ud \Omega_{\de}} = - \frac{\left( 1- \Omega_{\de} \right)^{\gamma}}{3 \Omega_{\de} \left[ \xi + w_{\de} \left(1 - \Omega_{\de} \right) \right] }.
    \end{gather}
    We integrate backwards in $\Omega_{\de}$ from $\Omega_{\de,0}$ to $\Omega_{\de}(z)$ and expand it to obtain
    \begin{align}
        \ln &\frac{\delta_{\dm}}{\delta_{\dm,0}} = \ln \left( \frac{\Omega_{\de}}{\Omega_{\de,0}} \right)^{-1/3 \tilde w_0 }
        + \frac{\gamma_0 - \bar\omega_{01} }{3 \tilde w_0} \left(\Omega_{\de} - \Omega_{\de,0} \right) - {} \nonumber \\
        {} &- \frac{1}{6 \tilde w_0}\left[\frac{\gamma_0^2}{2}  - \gamma_0  \left( \frac{1}{2} + \bar\omega_{01} \right) - \gamma_1 + \frac{1}{\tilde w_0} \left( w_0 \bar\omega_{01} - w_2  + \frac{w_1 \tilde w_1}{\tilde w_0} \right) \right] \left( \Omega_{\de}^2 - \Omega_{\de,0}^2 \right) + {}  \nonumber \\
        \label{eq:sigma8}
        {} &+ \mathcal{O}(\Omega_{\de}^3) + \mathcal{O}(\Omega_{\de,0}^3)
    \end{align}
    where we have introduced the definitions
    \begin{gather}
        \tilde w_n \equiv w_n + \xi \qquad \text{and} \qquad
        \bar\omega_{01} \equiv \frac{w_0 - w_1}{\tilde w_0}.
    \end{gather}
    The time dependence of $\delta_{\dm}$ is parametrized by $\Omega_{\de}$. $\delta_{\dm,0}$ and $\Omega_{\de,0}$ represent their values today.
    Eq.~\eqref{eq:sigma8} then gives
    \begin{align}
        \label{eq:sigma8m2}
        \delta_{\dm}(z) &= \delta_{\dm,0} \mathcal{D}(z;0), \quad \text{with} \quad \mathcal{D}(z;0) \equiv \left[ \frac{\Omega_{\de(z)}}{\Omega_{\de,0}} \right]^{-1/3 \tilde w_0} \exp \left[ \frac{\varepsilon_1 \Delta_{\de}^{(1)} + \varepsilon_2 \Delta_{\de}^{(2)}}{3 \tilde w_0} \right]
    \end{align}
    up to the second order in $\Omega_{\de}$ and $\Omega_{\de,0}$, with
    \begin{align}
        \varepsilon_1 &\equiv \gamma_0 - \bar\omega_{01}, \\
        \varepsilon_2 &\equiv -\frac{\gamma_0^{2}}{4} + \frac{\gamma_0}{2} \left(\frac{1}{2} + \bar \omega_{01} \right) + \frac{\gamma_1}{2} - \frac{1}{2 \tilde w_0} \left( w_0 \bar\omega_{01} - w_2 + \frac{w_1 \tilde w_1}{\tilde w_0} \right), \\
        \Delta_{\de}^{(n)} &\equiv \Omega_{\de}^n(z) - \Omega_{\de,0}^n.
    \end{align}
    $\mathcal{D}(z;0)$ is the backward propagation function for the evolution of the \gls{dm} perturbation from redshift zero to $z$.
    Noting that $P(k, z) = \left[ \mathcal{D}(z;0) \right]^2 P_0(k)$ and $\mathcal{D}$ is scale-independent, it follows directly from the definition \eqref{eq:s8def} that $\sigma_R^2 = \mathcal{D}^2 \sigma_{R,0}^2$, i.e., $\sigma_R$ satisfies the same equation \eqref{eq:sigma8m2} for $\delta_{\dm}$.
    Thus, at the scale $R = 8\,h^{-1}\,\si{\mega\parsec}$, we have
    \begin{align}
        \label{eq:normsigma8}
        \sigma_{8}(z) &= \sigma_{8,0} \left[ \frac{\Omega_{\de}(z)}{\Omega_{\de,0}} \right]^{-1/3 \tilde w_0} \exp \left[ \frac{\varepsilon_1 \Delta_{\de}^{(1)} + \varepsilon_2 \Delta_{\de}^{(2)}}{3 \tilde w_0} \right],
    \end{align}
    also up to the second order in $\Omega_{\de}$ and $\Omega_{\de,0}$.
    Note that there can be some inaccuracy in the computation of $\sigma_8(z)$ from eq.~\eqref{eq:normsigma8}, since we are integrating a function that has been expanded around $\Omega_{\de}=0$ from redshift zero, where $\Omega_{\de}$ is not so small, until $z$.
    This has the consequence of the errors of the expansion at low redshifts being accumulated for $\sigma_8$ at any redshift and constitutes a limitation of the method.
    We also note that if $|w_1|$ or $|w_2|$ is too large, it is possible that they can make the exponential in eq.~\eqref{eq:normsigma8} grow enormously.

    For the evaluation of $\Omega_{\de}(z)$, we have to use a recursive relation.
    The \gls{dm} and \gls{de} densities, in terms of the redshift, are 
    \begin{subequations}
        \begin{align}
            \bar\rho_{\dm}(z) &= \bar\rho_{\dm,0} \exp \left[ \int_0^z \frac{3}{1+\tilde z} \left(1 - \xi \frac{\Omega_{\de}}{1 - \Omega_{\de}} \right) \ud \tilde z \right], \\
            \label{eq:consde}
            \bar \rho_{\de}(z) &= \bar\rho_{\de,0} \exp \left[ \int_0^z \frac{3}{1+\tilde z}\left[1 + w_{\de}(\tilde z) + \xi \right] \ud \tilde z \right]. 
        \end{align}
    \end{subequations}
    The zero-th order \gls{de} density parameter is obtained by setting $w_{\de} = w_0$ and neglecting the term $\xi \frac{\Omega_{\de}}{1 - \Omega_{\de}} \approx \xi \Omega_{\de} + \xi \Omega_{\de}^2$,
    \begin{align}
        \Omega_{\de}^{(0)} = \frac{\bar\rho_{\de}^{(0)}}{\bar\rho_{\de}^{(0)} + \bar\rho_{\dm}^{(0)}} = \frac{\Omega_{\de,0} \left( 1 + z \right)^{3 \tilde w_0}}{1 - \Omega_{\de,0} + \Omega_{\de,0} \left(1 + z \right)^{3 \tilde w_0}}.
    \end{align}
    Now the density parameter up to the first order is calculated by using $w_{\de} = w_0 + w_1 \Omega_{\de}^{(0)}$ and $\xi \frac{\Omega_{\de}}{1 - \Omega_{\de}} = \xi \Omega_{\de}^{(0)}$,
    \begin{align}
        \label{eq:Omega1m2}
        \Omega_{\de}^{(1)}(z) = \frac{\Omega_{\de,0} \left(1 + z \right)^{3 \tilde w_0} \left[ 1 - \Omega_{\de,0} + \Omega_{\de,0} \left(1 + z \right)^{3 \tilde w_0} \right]^{\tilde w_1 / \tilde w_0}}{1 - \Omega_{\de,0} + \Omega_{\de,0} \left(1 + z \right)^{3 \tilde w_0} \left[1 - \Omega_{\de,0} + \Omega_{\de,0} \left(1 + z \right)^{3 \tilde w_0} \right]^{\tilde w_1 / \tilde w_0}}
    \end{align}
    With equations \eqref{eq:gammas}, \eqref{eq:normsigma8} and
    \eqref{eq:Omega1m2} we are now able to compute $f(z)$ and $\sigma_8(z)$
    provided that we know the parameters $\xi$, $w_n$, and $\Omega_{\de,0}$.
    Once we know the coupling, \gls{de} \gls{eos} coefficients, \gls{de} density
    parameter and the mean perturbation amplitude at present we can determine
    analytically how structures have evolved and can compare these results with
    \gls{lss} observations.

    \subsection{Stability conditions}
    Interacting \gls{de} models with constant \gls{eos} have already been shown to suffer from instabilities with respect to curvature and dark energy perturbations \cite{Valiviita2008,He2009139}.
    Depending on some combinations of the sign of the interaction and on the dark energy being of the quintessence or phantom type, $\delta_{\de}$ and the potential $\phi$ can blow up.
    Table~\ref{tab:instabilities} summarizes the allowed regions for the interaction and the \gls{de} equation of state parameters in the \gls{cde} model as shown by ref.~\cite{Gavela2009}, which extends the model stability analysis of ref.~\cite{He2009139} to negative values of $\xi$.
    \begin{table}[t]
        \setlength\tabcolsep{49pt}
        \caption{Stability conditions of the \gls{cde} model.}
        \label{tab:instabilities}
        \centering
        \begin{tabular}{@{}lcc@{}}
            \toprule
            Constant \gls{eos}  &   Interaction sign  &   Condition  \\
            \midrule
            $w_{\de} < -1$    &   $\xi < 0$   &   early-time instability   \\
            $w_{\de} < -1$    &   $\xi > 0$   &   stable      \\
            $-1 < w_{\de} < 0$    &   $\xi < 0$   &   stable  \\
            $-1 < w_{\de} < 0$    &   $\xi > 0$   &   early-time instability \\
            \bottomrule
        \end{tabular}
    \end{table}

    These results strongly restrict the parameter space for interacting \gls{de}.
    As those references point out, such instabilities can be avoided by allowing the \gls{eos} to vary with time, which we do when we expand $w_{\de}$ in terms of $\Omega_{\de}$.
    However, before considering a time variable \gls{eos}, first we simplify our models by fixing $w_1$ so we have one less parameter to be constrained with the \gls{mcmc} method.
    We proceed in the next section to compare our results for the growth rate with numerical calculations provided by a modified version of \texttt{CAMB} \cite{LewisCAMB}, in order to assess the reliability of our expressions and validate the method.

    \subsection{Comparison with full numerical computations in \texttt{CAMB}}
    \label{ss:CAMBcomp}
    To test how effective our analytical result of the growth in the \gls{cde} model is, we compare it with the numerical $f(z)$ obtained in a modified version of \texttt{CAMB} for the interacting model.\footnote{The model implemented in \texttt{CAMB} had a baryonic component that could be set to account for a minimum of \SI{0.2}{\percent} of the total energy density, which should be perfectly fine as our tests showed that even the higher amount of \SI{4}{\percent} did not have a perceptible influence on the comparisons. This modified version of \texttt{CAMB} has been used in previous works \cite{AndreElcio2014PRD89,AndreXiaodong2016}. See ref.~\cite{AndreThesis} for more details about the implementation.}
    We are going to show that our analytic solution can be trusted and we can further use it
    to estimate the cosmological parameters with a \gls{mcmc} code, as a shortcut alternative to the full numerical computation to speed up the calculation.

\begin{figure}[tb]
        \centering
        \includegraphics[width=\textwidth]{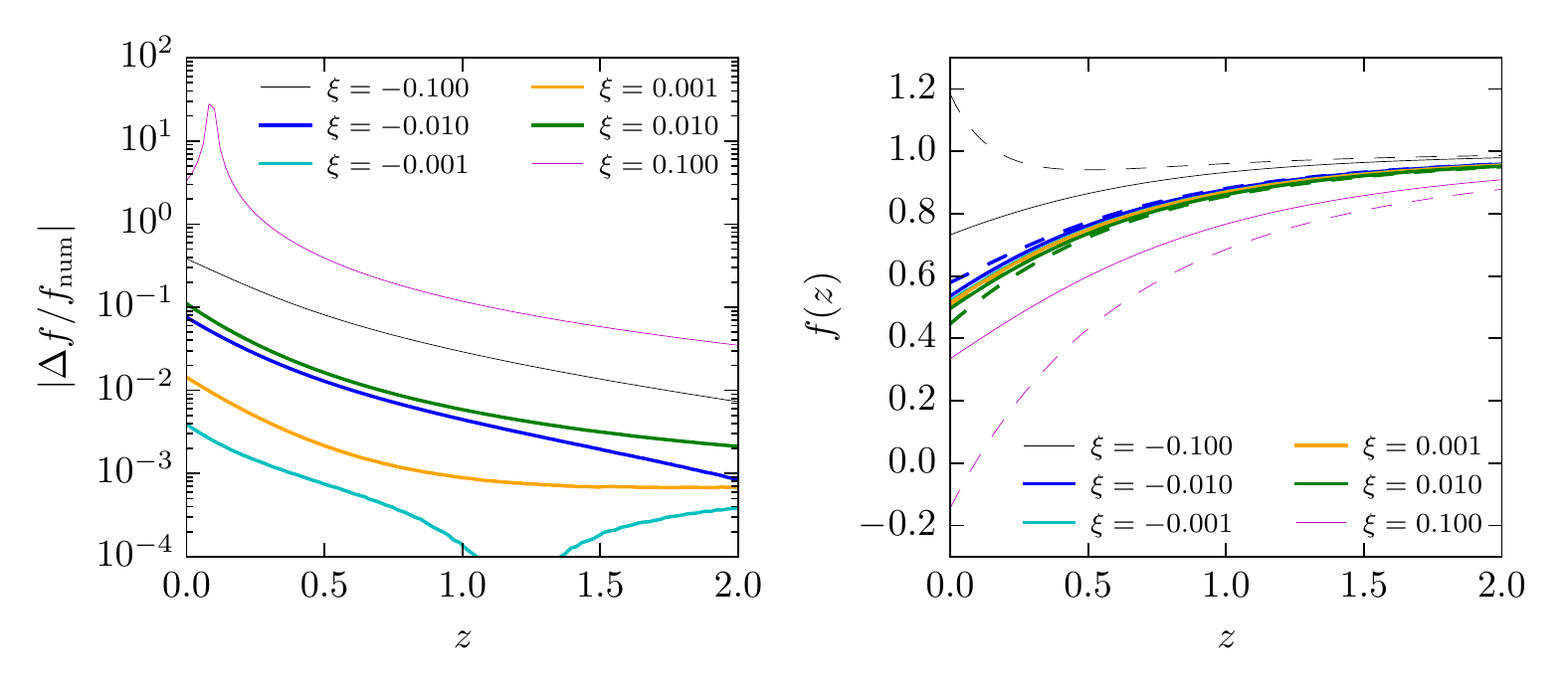}
        \caption{Comparison between analytical and numerical computations of $f(z)$ for the \gls{cde} model. In the left panel, the modulus of the relative differences, in logarithmic scale; in the right panel, the dashed lines represent the numerical results for $f(z)$, while the solid lines show our analytical results. Values of $|\xi|$ as big as $0.1$ give large discrepancies and should be avoided. We use thin lines to represent them.}
        \label{fig:compModel2}
    \end{figure}

    We fix $w_1 = 0$ and $\Omega_{\de,0} = 0.7$ and calculate $f(z)$ with $z$ ranging from $0$ to $10$.
    According to the stability conditions given in the last section, the interaction constant in \gls{cde} can be negative, in which case the dark energy \gls{eos} must be of quintessence type, or the coupling be positive with phantom type \gls{de} \gls{eos}.
    We then fix $w_0 = -0.999$ and test the interaction constants $\xi = -0.1, -0.01, -0.001$ and $w_0 = -1.001$ with $\xi = 0.001, 0.01, 0.1$. 
    To distinguish these two tests, we use CPDE and CQDE for phantom- and quintessence-type dark energy, respectively.
    The comparisons are shown in figure~\ref{fig:compModel2} through the modulus of the difference $\Delta f \equiv f_{\anl} - f_{\tnum}$ divided by $f_{\tnum}$ (left panel), where ``anl'' and ``num'' stand for analytical and numerical computations.
    Over the range of the \gls{lss} data (low redshift until $z \sim 1$), for a given $\xi$, the discrepancy grows as we approach $z = 0$, which is expected from the fact that $\Omega_{\de,0}$ is as big as $0.7$.
    The discrepancy tends to decrease as $z$ increases, but only until a certain redshift, when it can start to grow, albeit slowly.
    In the plot, we focus on the redshift range $\left[0, 2 \right]$.
    In the $\xi = 0.1$ case, $f_{\tnum}$ can become negative and the discrepancy is huge.
    This occurs because as $z$ decreases, $\frac{1-\Omega_{\dm}}{\Omega_{\dm}}$ increases and the second term inside the curly brackets in eq.~\eqref{eq:ddotdeltaModel2} dominates the coefficient of $\delta_{\dm}$ and changes its sign, leading to a negative growth.
    The analytical parametrization $(\Omega_{\dm})^{\gamma}$, on the other hand, obviously can never become negative.
    The numerical result for $\xi = -0.1$ shows that $f$ grows very rapidly at small redshifts as $z$ goes to zero, a behavior that is opposite to the other cases.
    This is due to a change of sign in the coefficient of $\dot\delta_{\dm}$ (Hubble drag) in eq.~\eqref{eq:ddotdeltaModel2}.
    We discard the cases $\xi = - 0.1$ and $\xi = 0.1$ as they are not well described by eq.~\eqref{eq:fOmegagamma}
    and restrict $\xi$ within the interval $\left[ -0.01, 0 \right]$ for the CQDE model and $\left[0, 0.01\right]$ for the CPDE model,
    which allow the difference between the numerical and the analytical results to be kept below about \SI{10}{\percent} 
    (with the other parameters fixed at reasonable values).
    Cusps indicating a change of sign of $\Delta f$ are observed in the curves of $|\Delta f/f_{\tnum}|$
    at higher redshifts (not shown).
    The fact that the analytical and numerical curves cross themselves instead of converging to a common plateau, with $f_{\tnum}$ becoming smaller than $f_{\anl}$ as $z$ becomes larger, might indicate some contribution of a decaying mode of the perturbation, which is out of the scope of this work.

    The conclusion is that the \gls{mcmc} analysis can be made with high efficiency using the analytic expressions --- especially useful when the computational power available is limited --- derived for $f$ in the interacting \gls{de} model, provided the parameters are restricted to the region where the discrepancy with respect to the numerical reference from \texttt{CAMB} is reasonably small.
    In the next section we present the \gls{rsd} data that we use to estimate the parameters of our interacting models via \gls{mcmc}.

    \section{Observational constraints}
    \label{sec:data}

    In this section we present the dataset used to constrain the parameters of our models. Because of the way those data were obtained, an adjustment to our growth rate $f$, calculated in a universe where \gls{dm} interacts with \gls{de}, is required before comparing with $f \sigma_8$ data.
    We explain in detail how the comparison must be made, then describe the statistical method employed in the analysis and discuss the results.

    \subsection{The data}
    One way of measuring growth of structure is through the effect of redshift-space distortions.
    Kaiser~\cite{Kaiser_1987} showed that the galaxy power spectrum $P^{\s}$ observed in redshift space is expected to be amplified with respect to the real power spectrum $P(k)$ by a factor that depends on the growth rate and on the cosine of the angle between the movement of the galaxies and the observation $\mu_{kz} \equiv \frac{1}{k} \mathbf{k} \cdot \mathbf{\hat{z}}$.
    The linear theory, with the plane-parallel approximation for distant observer, imposes the relation 
    \begin{align}
        P^{\s}(\mathbf{k}) = \left(1 + \beta \mu_{kz}^2 \right)^2 P(k),
    \end{align}
    where $\mathbf{k}$ is the wavevector, $k$ its modulus and $\mathbf{\hat{z}}$ the line-of-sight direction. $\beta$ is the so-called redshift-space distortion parameter, defined as $\beta \equiv f(z)/b(z)$, where $b(z)$ is a bias parameter relating the galaxy and matter density contrasts by $\delta_{\g} = b \delta_{\m}$.
    The galaxy overdensity is extracted from a galaxy redshift survey.
    The bias can be estimated as $\frac{\sigma_{8,\g}}{\sigma_{8,\m}}$, the ratio of \gls{rms} fluctuations of the two overdensity fields. 
    Multipole analysis of the anisotropy of the redshift-space power spectrum or correlation function in the redshift survey allows the observational determination of $\beta$.
    Thus, one gets the measurement $\beta \sigma_{8,\g} = f \sigma_{8,\m}$ of the growth of structure.
    The advantage of using $f \sigma_8$ rather than just $f$ to compare with model predictions is that the estimator $\beta \sigma_{8,\g}$ does not require the assumption of a bias model.
    Also, the determination of $\beta$ is affected only weakly by changes in the cosmology (through the determination of distances) \cite{SongPercival2009,Basilakos2013}.

    In table~\ref{tab:data} we list measurements of growth rate with their errors for various redshifts from different surveys like 2dF, 6dF, SDSS, BOSS and WiggleZ.
    \begin{table}[t]
        \setlength\tabcolsep{16.5pt}
        \caption{Observed growth rate data and their respective references.}
        \label{tab:data}
        \centering
        \sisetup{table-format=1.3(1)}
        \begin{tabular}{@{}S[table-format=1.3]SccS[table-format=1.2]Sc@{}}
            \toprule
            {$z$} &   {$f \sigma_8(z)$} &   Ref.  & &  {$z$} &   {$f \sigma_8(z)$}  &   Ref.  \\
            \midrule
            0.02   &    0.360   \pm   0.040 &   \cite{HudsonTurnbull2012} & &   0.40   &    0.419   \pm   0.041 &   \cite{Tojeiroetal2012}\\
            0.067   &   0.423   \pm   0.055 &   \cite{Beutleretal2012} & &  0.41   &    0.450   \pm   0.040 &   \cite{Blakeetal2011}\\
            0.10   &    0.37   \pm    0.13  &   \cite{Feix2015} & &  0.50   &    0.427   \pm   0.043 &   \cite{Tojeiroetal2012}\\
            0.17   &    0.510   \pm   0.060 &   \cite{SongPercival2009,Percivaletal2004}  & &   0.57   &    0.427   \pm   0.066 &   \cite{Reidetal2012}\\
            0.22   &    0.420   \pm   0.070 &    \cite{Blakeetal2011} & &   0.60   &    0.430   \pm   0.040 &   \cite{Blakeetal2011}\\
            0.25   &    0.351   \pm   0.058 &   \cite{Samushia01032012} & &  0.60   &    0.433   \pm   0.067 &   \cite{Tojeiroetal2012}\\
            0.30   &    0.407   \pm   0.055 &   \cite{Tojeiroetal2012} & &  0.77   &    0.490   \pm   0.180 &   \cite{Guzzo:2008ac,SongPercival2009}\\
            0.35   &    0.440   \pm   0.050 &   \cite{SongPercival2009,Tegmark2006} &  &  0.78   &    0.380   \pm   0.040 &   \cite{Blakeetal2011}\\
            0.37   &    0.460   \pm   0.038 &   \cite{Samushia01032012} & &  0.80   &    0.47    \pm   0.08 &   \cite{delaTorreetal2013}\\
            \bottomrule
        \end{tabular}
    \end{table}
    Most of those data are measured using \gls{rsd} and others are based on direct measurements of peculiar velocities \cite{HudsonTurnbull2012,Turnbulletal2012,Davisetal2011} or galaxy luminosities \cite{Feix2015}.

    \subsubsection{Corrections to the growth rate due to the altered continuity equation}
    In a standard cosmology, the coherent motion of galaxies is connected to the growth rate through the galaxy continuity equation $\theta_{\g} = - \mathcal{H} \beta \delta_{\g}$, built upon the matter continuity equation $\theta_{\m} = - \mathcal{H} f_{\m} \delta_{\m}$ with the density bias assumption $\delta_{\g} = b \delta_{\m}$ and without any bias for the velocities ($\theta_{\g} = \theta_{\m}$).
    Whether the \gls{rsd} parameter $\beta$ is measured from the power spectrum or from peculiar velocities, the $f \sigma_8$ data are based on the correspondence between $f/b$ and the velocity divergence as established by the continuity equation.
    When an interacting matter component is involved, these continuity equations do not hold anymore.
    We will now see what quantity corresponds to the velocity divergence $\theta_{\g}$ in an interacting \gls{de} model.

    We need to start over from the baryons and \gls{dm} continuity equations in the interacting model to write a continuity equation for matter on which a continuity equation for galaxies can be based.
    The two matter fluids now behave differently, one coupled to the dark energy fluid and the other uncoupled.
    For baryons, we still have $\dot \delta_{\ba} + \theta_{\ba} = 0$.
    With ${Q_0}^{\dm} = - 3 \mathcal{H} \xi \bar \rho_{\de}$, the \gls{dm} continuity equation was obtained in section~\ref{ss:pertequations},
    \begin{align}
        \dot \delta_{\dm} + 3 \mathcal{H} \xi \frac{\bar \rho_{\de}}{\bar \rho_{\dm}} \delta_{\dm} + \theta_{\dm} = 0.
    \end{align}
    Since the matter density $\rho_{\m}$ is the sum of the densities $\rho_{\ba}$ and $\rho_{\dm}$, the matter perturbation is $\delta_{\m} = \left( \bar \rho_{\ba} \delta_{\ba} + \bar \rho_{\dm} \delta_{\dm} \right) / \bar \rho_{\m}$ and its time derivative is
    \begin{align}
        \dot \delta_{\m} &= - 3 \mathcal{H} \xi \frac{\bar \rho_{\de}}{\bar \rho_{\m}} \delta_{\m} - \frac{\bar \rho_{\ba} \theta_{\ba} + \bar \rho_{\dm} \theta_{\dm}}{\bar \rho_{\m}}
    \end{align}
    where we have also used the background evolution equations of each component from eq.~\eqref{eq:Bianchi0background}.
    Substituting the time derivative,
    \begin{align}
        \mathcal{H} \left( \frac{\ud \ln \delta_{\m}}{\ud \ln a} + 3 \xi \frac{\bar \rho_{\de}}{\bar \rho_{\m}} \right) \delta_{\m}  + \frac{\bar \rho_{\ba} \theta_{\ba} + \bar \rho_{\dm} \theta_{\dm}}{\bar \rho_{\m}} = 0.
    \end{align}
    Recognizing $\theta_{\m}$ by the term $\left( \bar\rho_{\ba} \theta_{\ba} + \bar \rho_{\dm} \theta_{\dm} \right)/ \bar \rho_{\m}$, as usual, gives the continuity equation altered by the interaction
    \begin{align}
        \mathcal{H} \tilde f_{\m} \delta_{\m} + \theta_{\m} = 0,
    \end{align}
    where $\tilde f_{\m} \equiv f_{\m} + 3 \xi \frac{\bar \rho_{\de}}{\bar \rho_{\m}}$ is the modified growth rate, with the usual $f_{\m} \equiv \frac{\ud \ln \delta_{\m}}{\ud \ln a}$.

    We maintain the assumption that galaxies trace the matter field via $\delta_{\g} = b \delta_{\m}$ and $\theta_{\g} = \theta_{\m} = \theta$, so the galaxy continuity equation is now
    \begin{align}
        \label{eq:modbeta}
        \mathcal{H} \tilde \beta \delta_{\g} + \theta = 0,
    \end{align}
    with $\tilde \beta \equiv \tilde f/ b$.
    Therefore, this modified growth rate function is the quantity that effectively corresponds to the coherent motion of galaxies if there is an interaction between \gls{dm} and \gls{de} according to the \gls{cde} model considered here.
    Also, the \gls{rsd} parameter that is effectively measured from the power spectrum is $\tilde \beta$, since the modeling of the Kaiser effect, including its nonlinear features, relies on a continuity equation like eq.~\eqref{eq:modbeta}\footnote{Higher-order terms are generally neglected in the continuity equation.} to substitute the velocity divergence in favor of the density multiplied by the (thus modified) growth rate.
    The same argument applies about the treatment of nonlinear effects like the Fingers-of-God (FoG) (see, for example, refs.~\cite{OhSeon2016,Taruya2010}).
    We then just need to add the term $3 \xi \frac{\bar \rho_{\de}}{\bar \rho_{\m}}$ to the growth rate $f_{\dm} = \Omega_{\dm}^{\gamma}$ obtained in section~\ref{ss:CDEmodel} before comparing those predictions to the $f \sigma_8$ data.
    In our simplified model with the matter sector composed of dark matter only, without baryonic matter, the modified growth rate is $\tilde f_{\dm} = f_{\dm} + 3 \xi \frac{1 - \Omega_{\dm}}{\Omega_{\dm}}$.

    \subsection{The statistical method}
    We perform a posterior likelihood analysis with flat priors for the parameters. 
    In order to do that, we employ our analytic formula in computing the theoretical growth, implement it in a \gls{mcmc} program in python and carry out the data fitting by using a simple Metropolis algorithm \cite{Geyer_2011,emcee,Heavens2010}.
    The proposal function in the algorithm is a multivariate normal distribution centered at the current state of the Markov chain.
    Its covariance matrix is a diagonal matrix where each diagonal element is equal to the square of a fraction of the prior interval of its corresponding parameter, adjusted by hand to give an acceptance ratio roughly between $0.2$ and $0.5$ in the Metropolis algorithm \cite{emcee}. 
    The likelihoods are computed as $\log \mathcal{L} = - \sum_{i=1}^{N} \log\left(\sigma_i \sqrt{2\uppi}\right) - \chi^2/2$, with
    \begin{align}
        \chi^2 = \sum_{i=1}^{N} \frac{\left[f\sigma_8^{\text{\scriptsize (obs)}}(z_i) - \tilde f \sigma_8^{\text{\scriptsize (th)}}(z_i)\right]^2}{\sigma_i^2}.
    \end{align}
    $N$ is the number of points in the dataset, $\sigma_i$ the errors in the measurements, ``obs'' stands for the observed data and ``th'' is our theoretical prediction by using the analytic formula on the growth. We then compute the unnormalized posterior $P(X|D) \propto P(D|X) \pi(X)$ for the parameter-space point $X$ given the dataset $D$, according to the Bayesian theorem, where $P(D|X)$ is the likelihood $\mathcal{L}$ and $\pi(X)$ is the prior.
    Our \gls{mcmc} code evolves the chains checking for convergence after each $N_{\text{\scriptsize steps}}$ and keeps running until they match the convergence criteria.
    The starting points are chosen randomly with uniform probability within the prior ranges for each parameter.
    For monitoring the convergence of the chains, we implemented the multivariate extension of the method proposed by Gelman and Rubin \cite{Gelman_1992,Brooks_1998}.

    \subsection{The results}
    \label{ss:results}
\begin{figure}[t]
        \centering
        \includegraphics[width=0.6\textwidth]{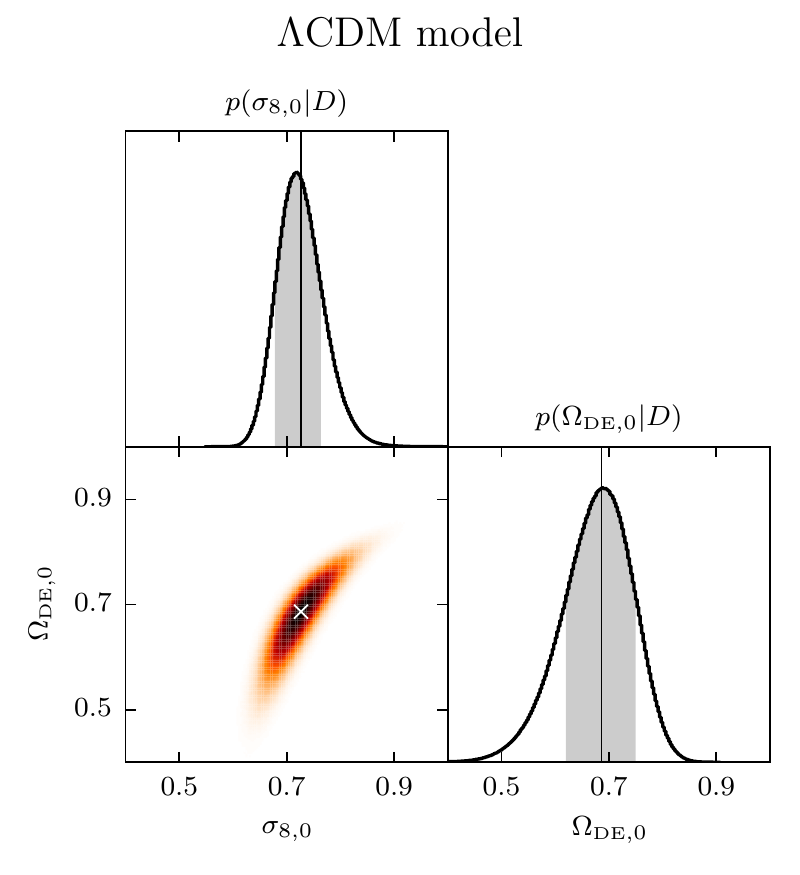}
        \caption{Histograms for the values of the \gls{de} density parameter and
            dark matter \gls{rms} fluctuation today at the scale of
            $8\,h^{-1}\,\si{\mega\parsec}$ in the $\Lambda$CDM model.  The
            vertical thin lines mark the best-fit values, and the grey area
            under the histograms show the $1\sigma$ \gls{cl}. In the 2D
            histogram, the colors map the parameter space points to their
            unnormalized posterior values, from white (lowest values) to black
            (highest values), with shades of orange representing intermediate values. The white cross marks the best-fit point.}
        \label{fig:LCDMgrid}
    \end{figure}

    For comparison purposes, we first constrain a simple $\Lambda$CDM model with the two free parameters $\sigma_{8,0}$ and $\Omega_{\de,0}$. Their best-fit values are used in the subsequent analysis when we compare the fitting of our models to the $\Lambda$CDM's fitting with the same data in section~\ref{sss:comparemodels}.
    The $f\sigma_8$ data from table~\ref{tab:data} provide the following $1\sigma$ \gls{cl} for the parameters: $\sigma_{8,0} = 0.7195^{+0.0440}_{-0.0415}$, $\Omega_{\de,0} = 0.6889^{+0.0606}_{-0.0691}$, with the best-fit values $\sigma_{8,0} = 0.7266$ and $\Omega_{\de,0} = 0.6864$ (see figure~\ref{fig:LCDMgrid}). 
    The priors used were $\left[ 0.4, 1.0 \right]$ for both parameters and we summarize the results in table~\ref{tab:allmodels}.
    \begin{table}[t]
        \centering
        \setlength{\tabcolsep}{18pt}
        \caption{Priors, best-fit values and $1\sigma$ \gls{cl} ranges for the parameters of all models. Central values are shown only for reasonably well constrained parameters.}
        \label{tab:allmodels}
        \sisetup{table-format=1.4}
        \begin{tabular}{@{}lccSc@{}}
            \toprule
            Model   &   Parameter   &   Prior   &   {Best-fit}    &   $1\sigma$ \gls{cl}  \\
            \midrule
            \multirow{2}{*}{$\Lambda$CDM}   &   $\sigma_{8,0}$	 & 	$\left[0.4, 1.0\right]$	 & 	0.7266	 & 	$0.7195^{+0.0440}_{-0.0415}$	\\
                                    &   $\Omega_{\de,0}$	 & 	$\left[0.4, 1.0\right]$	 & 	0.6864	 & 	$0.6889^{+0.0606}_{-0.0691}$	\\            
            \midrule
            \multirow{4}{*}{CPDE}   &   $\xi$	 & 	$\left[0.00, 0.01\right]$	 & 	\num{7.8e-5}	 & 	$\left[0.0034, 0.0100\right]$	\\
                                        &   $\sigma_{8,0}$	 & 	$\left[0.2, 1.4\right]$	 & 	0.6750	 & 	$0.6322^{+0.0473}_{-0.0293}$	\\
                                        &   $\Omega_{\de,0}$	 & 	$\left(0.0, 1.0\right]$	 & 	0.6712	 & 	$0.6939^{+0.0652}_{-0.0731}$	\\
                                        &   $w_0$	 & 	$\left[-3.0, -1.0\right)$	 & 	-1.4173	 & 	$\left[-2.1042, -1.0000\right]$	\\            
            \midrule
            \multirow{4}{*}{CQDE}   &   $\xi$	 & 	$\left[-0.01, 0.00\right]$	 & 	-0.0100	 & 	$\left[-0.0069, 0.0000\right]$	\\
                                        &   $\sigma_{8,0}$	 & 	$\left[0.2, 1.4\right]$	 & 	0.7230	 & 	$0.7513^{+0.1262}_{-0.0598}$	\\
                                        &   $\Omega_{\de,0}$	 & 	$\left(0.0, 1.0\right]$	 & 	0.6533	 & 	$0.7032^{+0.0667}_{-0.0705}$	\\
                                        &   $w_0$	 & 	$\left(-1.0, -0.3\right]$	 & 	-0.9977	 & 	$\left[-1.0000, -0.5552\right]$	\\            
            \midrule
            \multirow{3}{*}{$w$CQDE}    &   $\xi$	 & 	$\left[-0.01, 0.00\right]$	 & 	-0.0100	 & 	$\left[-0.0100, -0.0031\right]$	\\
                                            &   $\sigma_{8,0}$	 & 	$\left[0.2, 1.4\right]$	 & 	0.7240	 & 	$0.7166^{+0.0412}_{-0.0386}$	\\
                                            &   $\Omega_{\de,0}$	 & 	$\left(0.0, 1.0\right]$	 & 	0.6546	 & 	$0.6737^{+0.0512}_{-0.0702}$	\\
            \bottomrule
        \end{tabular}
    \end{table}    
    The growth rate determined by the \gls{eos} parameters is
    \begin{align}
        \text{$\Lambda$CDM:}    & \qquad f(\Omega_{\dm}) = \left( \Omega_{\dm} \right)^{0.5455 + 0.0073 \left(1 - \Omega_{\dm} \right)}
    \end{align}
    regardless of the resulting best-fit $\sigma_{8,0}$ and $\Omega_{\de,0}$. 
    The growth index today is $\gamma = 0.5505$, up to first order in $\Omega_{\de}$.
    In the following, we present the results for the interacting \gls{de} models.

    \subsubsection{The coupled \gls{de} models}
    \label{sss:model1}
\begin{figure}[t]
        \centering
        \subfloat[\label{fig:fdiffxi}Effect of $\xi$ on $\tilde f(z)$ (top) and $\sigma_8(z)$ (bottom panel).]{\includegraphics[width=0.47\textwidth]{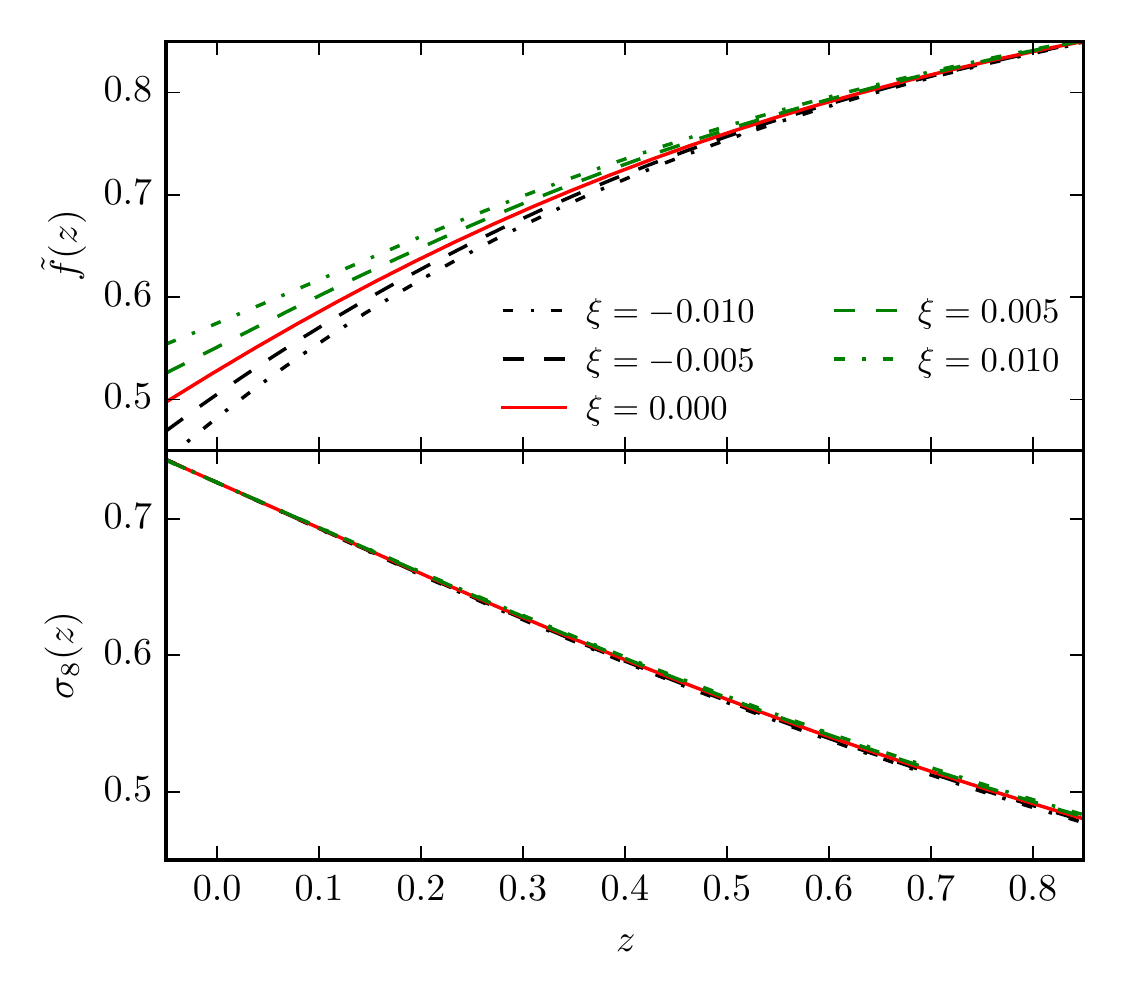}} \hspace{22pt}
        \subfloat[\label{fig:fs8diffxi}Influence of different values of $\xi$ on the product $\tilde f\sigma_{8}$.]{\includegraphics[width=0.47\textwidth]{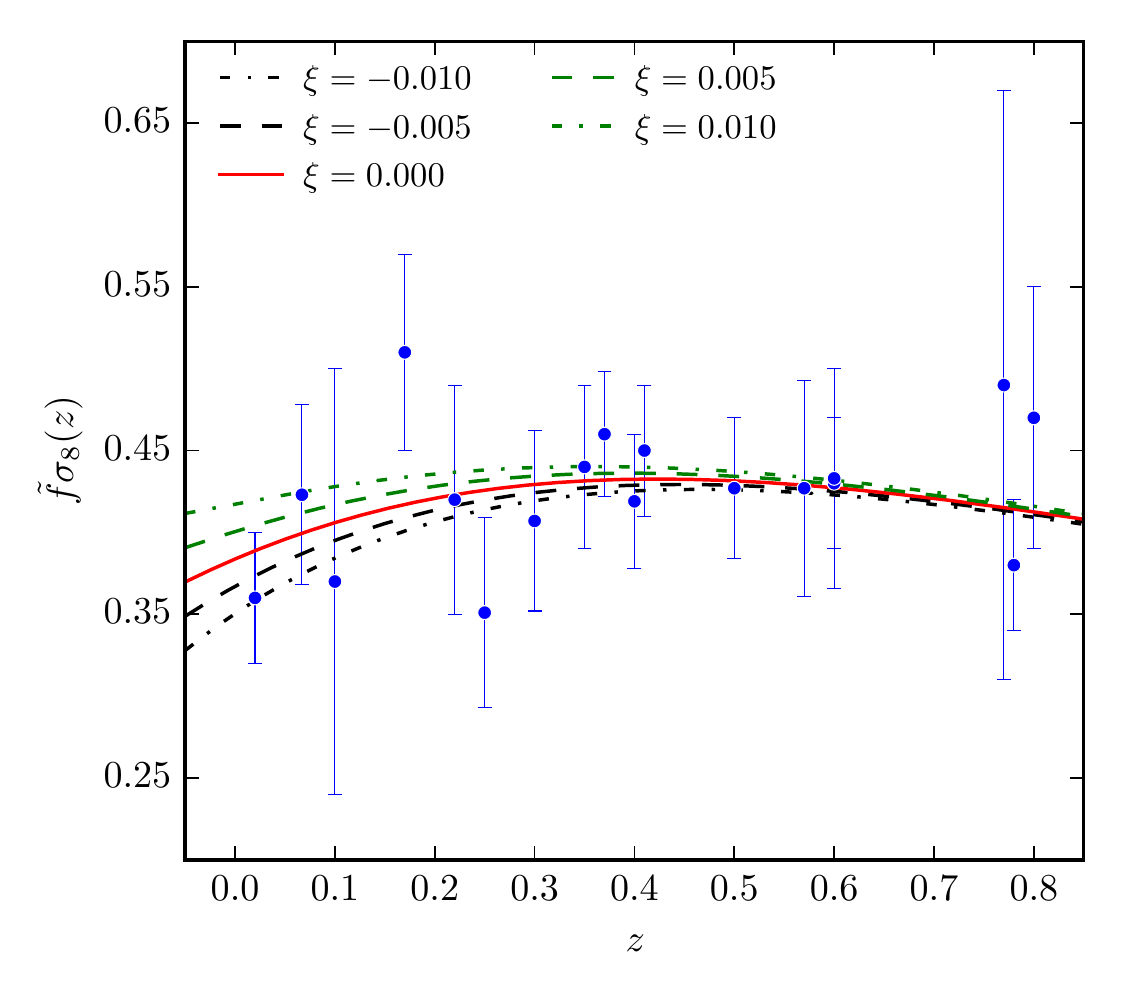}}
        \caption{Evolution of the growth of structures in the coupled \gls{de} model for varying values of the coupling $\xi$. The negative values (black lines) correspond to the CQDE model and the positive values (green lines) to the CPDE model. In both cases we use $w_0 = -1$ for simplification, since we are interested in seeing the effect of the coupling only. The red line is the $\Lambda$CDM result. The data from table~\ref{tab:data} are also plotted in (b).}
        \label{fig:differentxi}
    \end{figure}

    Besides $\Omega_{\de,0}$ and $\sigma_{8,0}$, \gls{cde} has other free parameters: $w_0$, $w_1$ and the coupling constant $\xi$.
    However, before trying to constrain all these parameters together, we first fix $w_1 = 0$ and see if we can have a good indication of $w_0 \ne -1$.
    Not being able to constrain $w_0$ alone in the equation of state means that we will certainly not be able to constrain $w_0$ and $w_1$ together.
    We show in figure~\ref{fig:differentxi} the effect of the interaction on $\tilde f(z)$, $\sigma_8(z)$ and on the product $\tilde f \sigma_8(z)$ with $\Omega_{\de,0}$ and $\sigma_{8,0}$ fixed at their $\Lambda$CDM best-fit values and with $w_0 \rightarrow -1$.
    In \ref{fig:fdiffxi} (top panel) we can clearly see influence of the interaction on the growth rate.
    The constant $\xi$ causes a shift of opposite sign to the growth rate $f$ (not shown), but a larger shift of equal sign to the modified rate $\tilde f$, the shift getting larger as $z$ gets closer to zero.
    The impact of the interaction on $\sigma_8$ (bottom panel) is barely perceptible.

    We choose the priors based on our comparison with the numerical result for $f(z)$, given in section~\ref{ss:CAMBcomp}.
    As discussed in section~\ref{ss:intromodel2}, in order to avoid changing the sign of the coefficient of $\delta_{\dm}$ and to keep discrepancies with respect to the numerical solutions small, values of $\xi$ should be small, of the order \num{e-2}, so we use the prior $\left[ 0 , 0.01 \right]$ for $\xi$ in the phantom case and  $\left[ -0.01, 0 \right]$ in the quintessence case.
    $\Omega_{\de,0}$ can be assumed any value in the interval $\left(0.0, 1.0 \right]$.

\begin{figure}[t]
        \centering
        \includegraphics[width=0.9\textwidth]{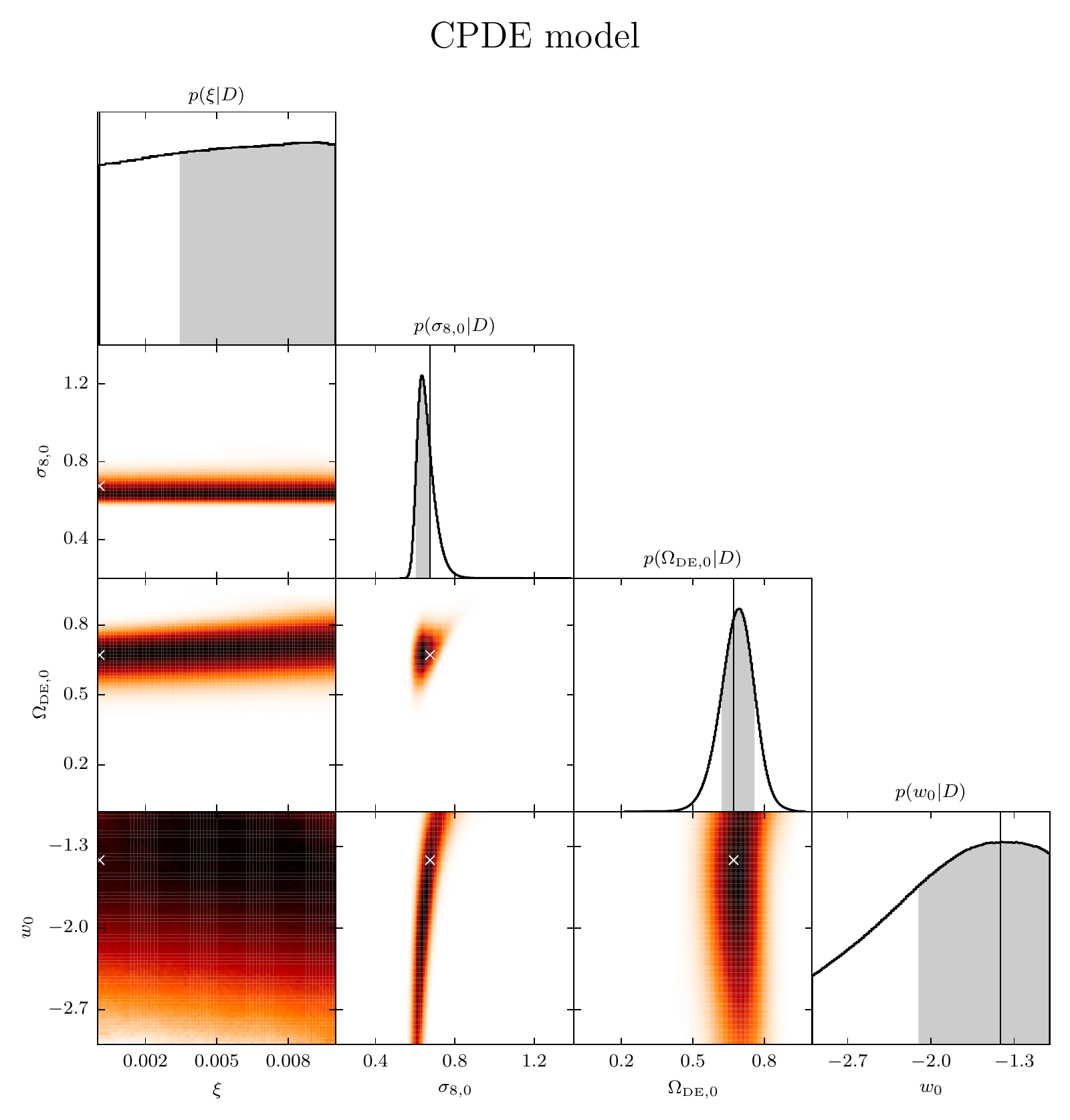}
        \caption{Histograms for the free parameters of CPDE. The vertical thin lines mark the best-fit values, and the grey area under the histograms show the $1\sigma$ \gls{cl}. In the 2D histograms, the colors map the parameter space points to their unnormalized posterior values, from white (lowest values) to black (highest values), with shades of orange representing intermediate values. The white crosses mark the best-fit point. Due to the large uncertainties in the \gls{rsd} measurements, the data could not constrain the interaction and the \gls{eos} parameter.}
        \label{fig:phantomgrid}
    \end{figure}

\begin{figure}[t]
        \captionsetup[subfigure]{position=top}
        \centering
        \subfloat[CQDE model]{\label{fig:model2Qgrid}\includegraphics[width=0.56\textwidth]{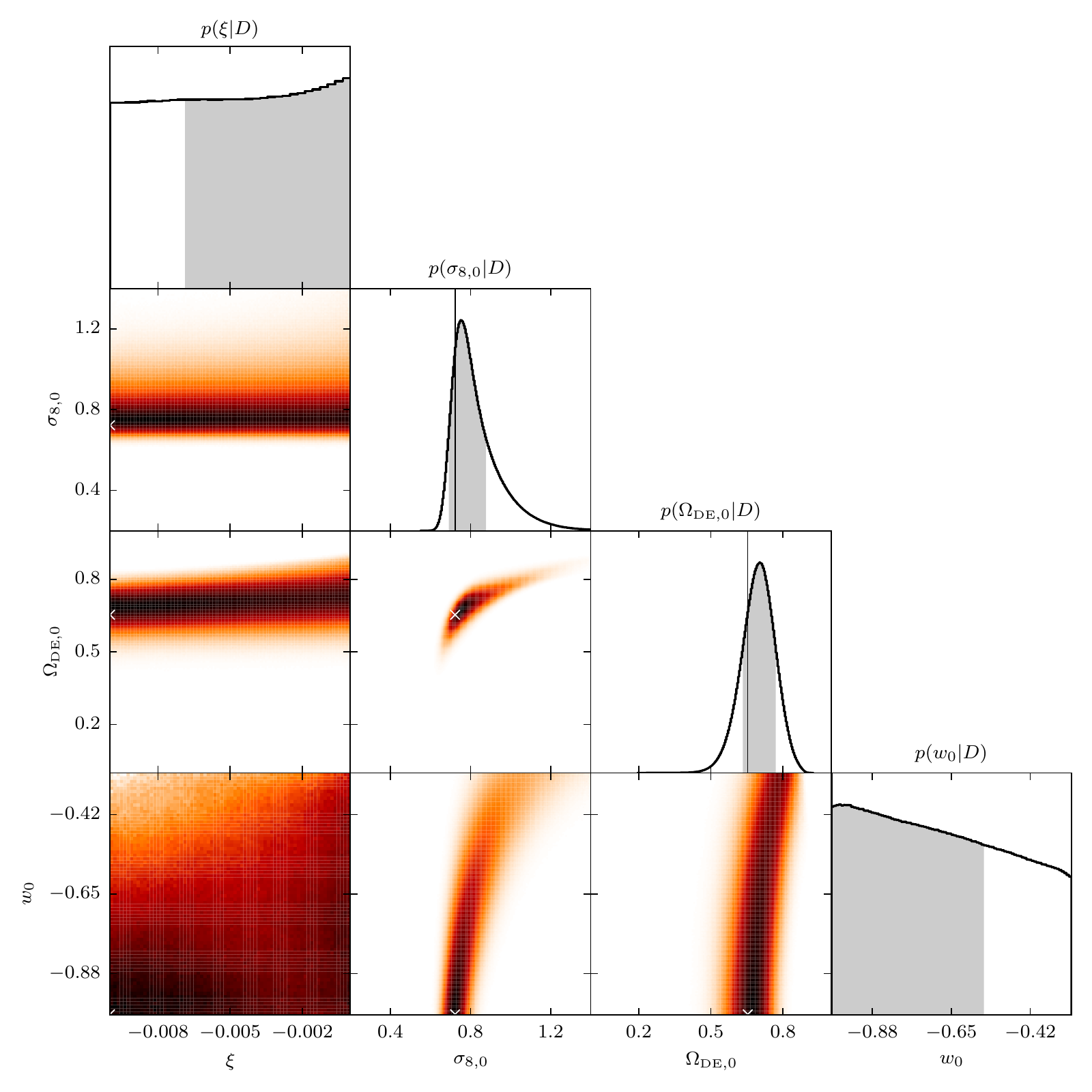}} 
        \subfloat[$w$CQDE model]{\label{fig:model2Lgrid}\includegraphics[width=0.43\textwidth]{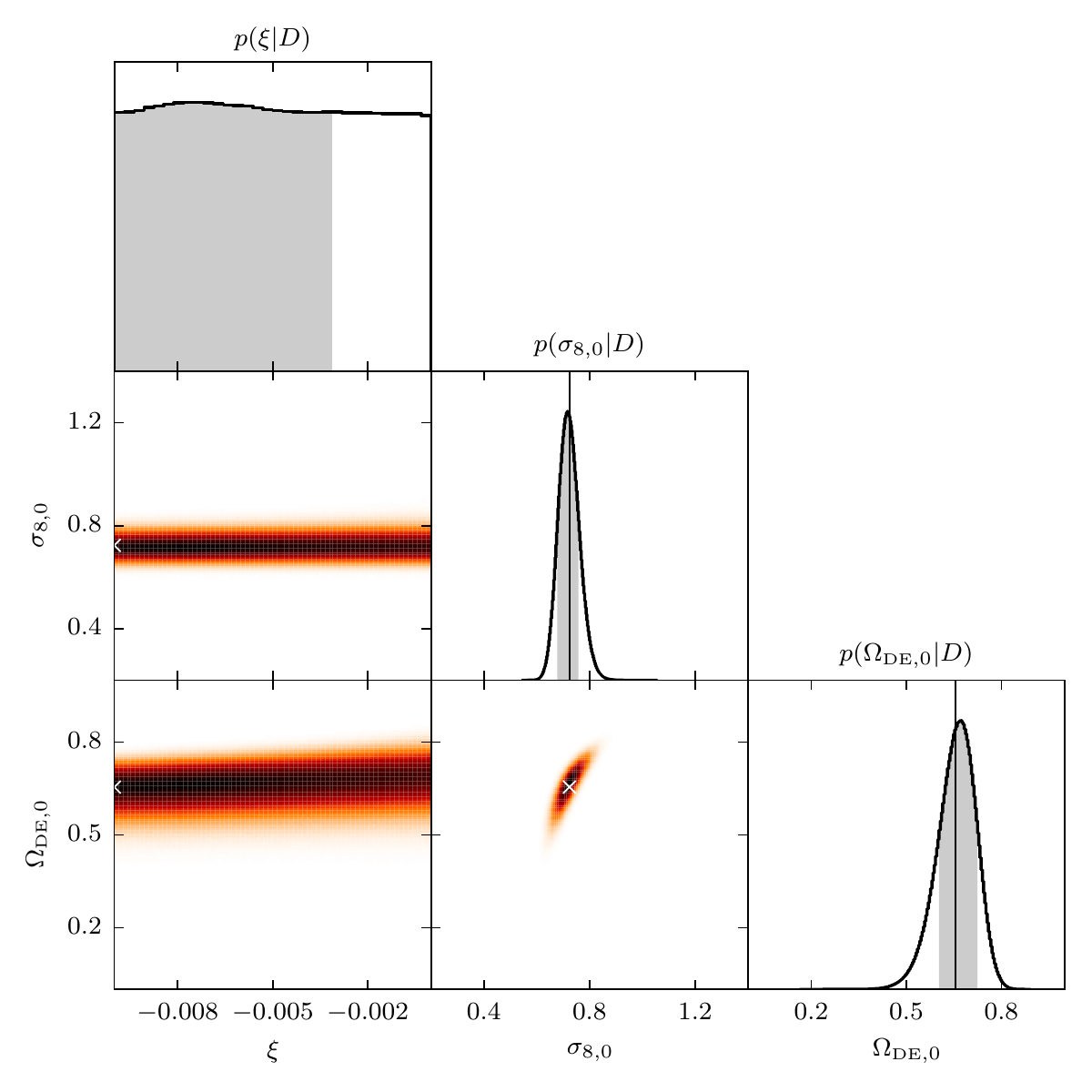}}
        \caption{Marginalized posterior distributions for (a) CQDE and (b) $w$CQDE models. The vertical thin lines mark the best-fit values, while the grey areas under the histograms in the diagonal show the $1\sigma$ \gls{cl}.  In the 2D histograms, the colors map the parameter space points to their unnormalized posterior values, from white (lowest values) to black (highest values), with shades of orange representing intermediate values. The white crosses mark the best-fit point. As we can see from the results of $w$CQDE, fixing the \gls{eos} parameter is not sufficient to constrain the interaction coupling in the already too tight prior.}
        \label{fig:quintessencegrid}
    \end{figure}

    Table~\ref{tab:allmodels} summarizes the priors and the fitting results and we show in figures~\ref{fig:phantomgrid} and \ref{fig:model2Qgrid} the marginalized distributions for CPDE and CQDE, respectively.
    We prefer to express the $1\sigma$ \gls{cl} intervals of the unconstrained parameters without reporting a central value.
    Because of the large uncertainties of the data, the method was not able to constrain $w_0$ and $\xi$ with $f\sigma_8$ data alone, as can be seen from the histograms of the marginalized distributions.
    This hints the fact that such set of parameters can only be better constrained if we combine the $f\sigma_8$ data with other kinds of observations, e.g.~the \gls{cmb}.
    The best-fit values encountered lead to the growth rates
    \begin{align}
        \text{CPDE:}     & \qquad f(\Omega_{\dm}) = \left( \Omega_{\dm} \right)^{0.5371 + 0.0058 \left( 1-  \Omega_{\dm} \right)}, \\
        \text{CQDE:}     & \qquad f(\Omega_{\dm}) = \left( \Omega_{\dm} \right)^{0.5290 - 0.0147 \left( 1-  \Omega_{\dm} \right)},
    \end{align}
    for the two models as functions of $\Omega_{\dm}$.
    The best-fit $\Omega_{\de,0}$ gives, for each model, the growth index today $\gamma = 0.5410$ and $\gamma = 0.5194$ respectively, up to first order in the density parameter.

    \subsubsection{On the unconstrained parameters}
    The models considered in our work cannot have all their parameters satisfactorily constrained due to the large uncertainties in the measurements of the large-scale structure.
    This difficulty motivated us to try to obtain a more conclusive determination of the interaction constant by fixing one more parameter, $w_0$ in the equation of state.
    We analyze the case of CQDE with the \gls{eos} fixed in its best-fit value $w_0 = -0.997728$.
    The choice of CQDE over CPDE is because this class of models gives, according to ref.~\cite{Gavela2009}, the best fit to \gls{lss} data.\footnote{Which model gives the best fit to the data that we used here could be evaluated by comparing their Bayesian evidences. However, this analysis is out of the scope of this work.}
    We then run this CQDE model with the \gls{eos} parameters fixed at $w_0 = -0.997728$ and $w_1 = 0$, which we call $w$CQDE. 
    The results are shown in figure~\ref{fig:model2Lgrid} and in table~\ref{tab:allmodels}.
    We obtained the growth rate
    \begin{align}
        \text{$w$CQDE:}   &  \qquad f(\Omega_{\dm}) = \left(\Omega_{\dm}\right)^{0.5290 - 0.0147 \left(1 - \Omega_{\dm}\right)},
    \end{align}
    with today's value of the growth index $\gamma = 0.5194$.
    This pretty much coincides with the CQDE result, since the best-fit values of all parameters are practically identical.

    We see that even when we fix the equation of state, although the region of $1\sigma$ \gls{cl} has been considerably reduced for $\sigma_{8,0}$ and $\Omega_{\de,0}$, the growth of structure data cannot constrain very well all parameters either because the measurements are not very precise or the prior is too tight.
    Relaxing this prior for $\xi$ would compromise the analysis, as the results for $f$ would not be so reliable, as discussed in section~\ref{ss:CAMBcomp}.
    This last result reinforces the need of additional observables in order to get fully satisfactory constraints and make assertive conclusions about a possible detection of a \gls{de}-\gls{dm} interaction.

    Indeed, Yang \& Xu \cite{YangXu2014apr} used \gls{cmb}, \gls{bao} and \gls{sneIa} in addition to $f\sigma_8$ data to constrain an interacting $w$CDM model (IwCDM) which is equivalent to our CQDE model.
    Murgia, Gariazzo and Fornengo \cite{Murgia2016} also combined \gls{cmb} temperature and polarization, gravitational lensing and supernovae data with \gls{bao}/\gls{rsd} data to constrain their models MOD1 and MOD2, identical to our models CQDE and CPDE, respectively.
    In ref.~\cite{AndreXiaodong2016}, the authors combined the latest Planck \gls{cmb} data, \gls{bao}, \gls{sneIa}, $H_0$ data and \gls{rsd} to constrain several parameters of their models, which also include our models CQDE and CPDE (models I and II in ref.~\cite{AndreXiaodong2016}).
    In all these works, the authors 
    obtained the growth by numerically computing the perturbation equations and compared with observational datasets.
    Their results are consistent with our treatment by employing the analytic formula on computing the growth.
    All these results converge that $f\sigma_8$ data alone cannot help to constrain well the model parameters due to the large uncertainty of the current data.

    \subsubsection{Comparing the growth in different models}
    \label{sss:comparemodels}
\begin{figure}[tb]
        \centering
        \includegraphics[width=\textwidth]{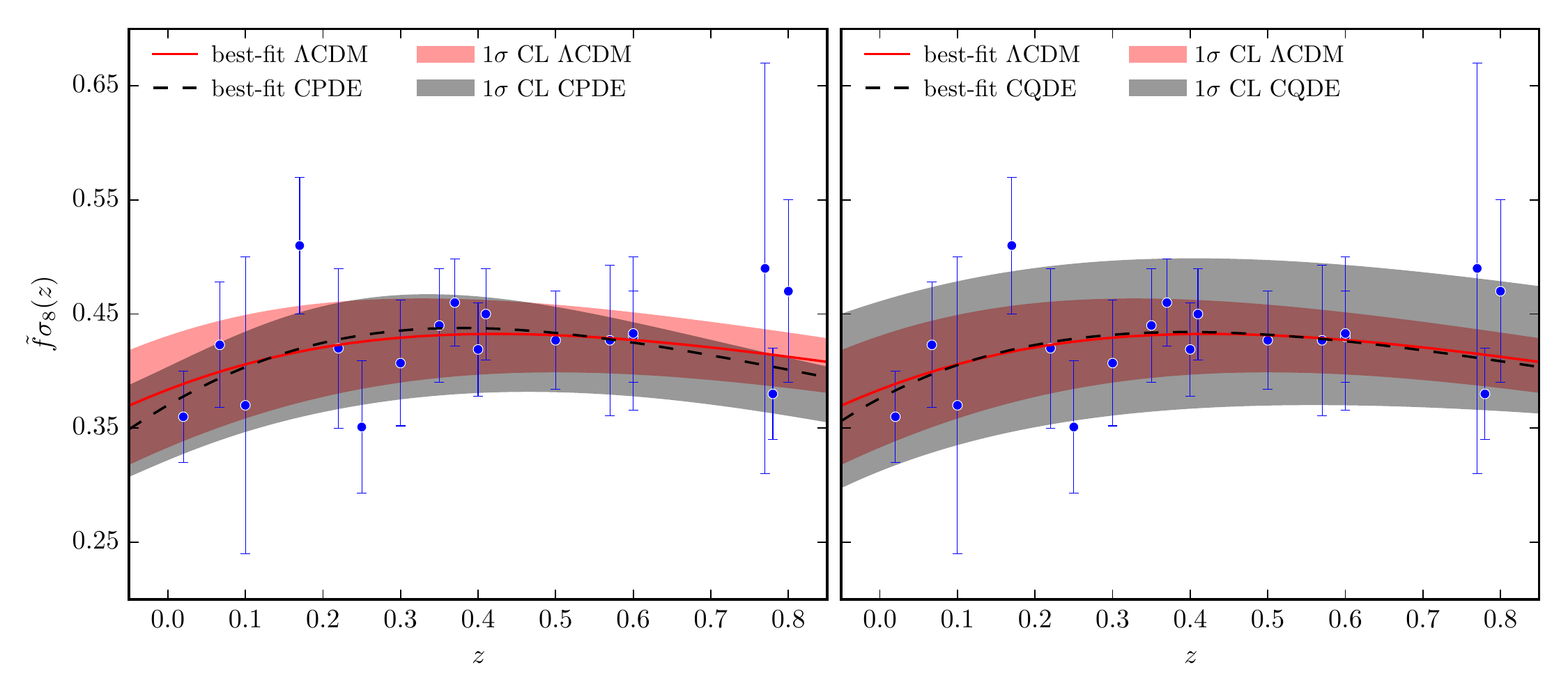}
        \caption{Comparisons of best-fit and $1\sigma$-range $\tilde f \sigma_8(z)$ between CPDE and $\Lambda$CDM (left panel) and between CQDE and $\Lambda$CDM (right panel). The blue data points are listed in table~\ref{tab:data}.}
        \label{fig:fs8onesigma}
    \end{figure}

    In figure~\ref{fig:fs8onesigma} we plot separately each of the interacting models' best-fit $\tilde f \sigma_8(z)$, together with the $\Lambda$CDM's best-fit over the redshift range of the data.
    We note that the best-fit $\tilde f \sigma_8(z)$ in the \gls{cde} models is generally lower than that in $\Lambda$CDM, but as the redshift decreases, it surpasses $\Lambda$CDM around redshift $z = 0.5$ and becomes smaller again around $z = 0.1$, the difference being slightly larger in the CPDE case (left panel) due to the best-fit point more distant from the $\Lambda$CDM best-fit.

    The discrepancies between the models become more apparent when we look at the $1\sigma$ ranges and at the functions $\tilde f(z)$, $\sigma_8(z)$ and $\gamma(z)$ separately.
    In order to do that, we perform linear error propagation on the fitted parameters.
    We simplify the task by centralizing the $1\sigma$ \gls{cl} intervals, getting the values listed in table~\ref{tab:1CLparams}, then propagate the errors through eqs.~\eqref{eq:gammas}, \eqref{eq:fOmegagamma}, \eqref{eq:normsigma8} and \eqref{eq:Omega1m2}.
    \begin{table}[t]
        \setlength\tabcolsep{24pt}
        \caption{Centralized $1\sigma$ \gls{cl} intervals of the free parameters in $\Lambda$CDM and interacting models for the linear error propagation.}
        \label{tab:1CLparams}
        \centering
        \sisetup{table-format=1.4(1)}
        \begin{tabular}{@{}cSSS@{}}
            \toprule
            {Parameter}    &   {$\Lambda$CDM}    &   {CPDE}    &   {CQDE}    \\
            \midrule
            $\xi \pm \Delta \xi$     & 0.0   &  0.0067 \pm 0.0033     &     -0.0034 \pm 0.0034  \\
            $\sigma_{8,0} \pm \Delta \sigma_{8,0}$     &  0.7209 \pm 0.0426   &  0.6412 \pm 0.0383     &     0.7845 \pm 0.0930  \\
            $\Omega_{\de,0} \pm \Delta \Omega_{\de,0}$    &   0.6846 \pm 0.0649 &   0.6900 \pm 0.0692 &   0.7013 \pm 0.0686 \\
            $w_0 \pm \Delta w_0$    &  -1.0 &   -1.5521 \pm 0.5521  &   -0.7776 \pm 0.2224    \\
            \bottomrule
        \end{tabular}
    \end{table}

    Although CQDE's best-fit is closer to $\Lambda$CDM than CPDE's best-fit, CQDE presents a wider $1\sigma$ range, encompassing the entire $\Lambda$CDM $1\sigma$ range (see figure~\ref{fig:fs8onesigma}).
    CPDE's $1\sigma$ range is about as wide as $\Lambda$CDM's.
    The three models are overall consistent within $1\sigma$ \gls{cl}.

\begin{figure}[tb]
        \centering
        \includegraphics[width=\textwidth]{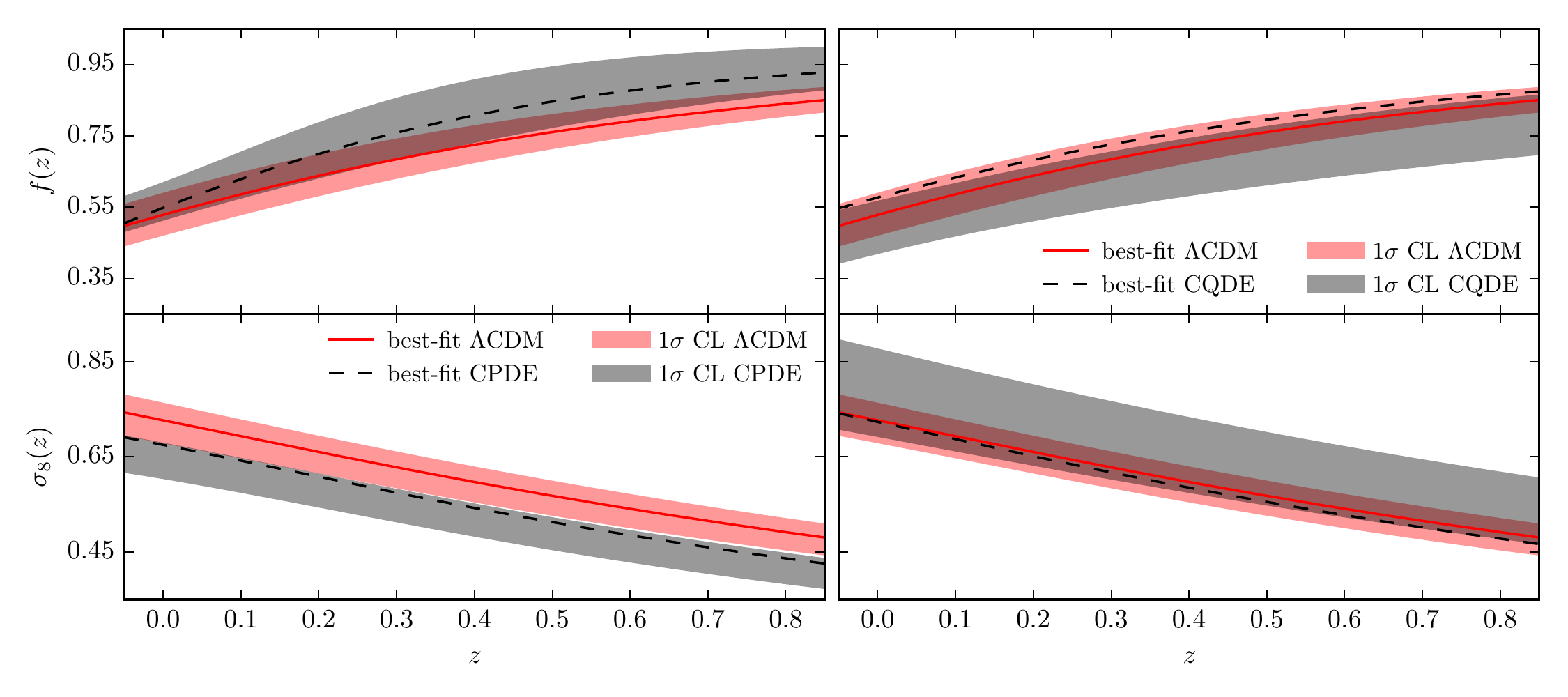}
        \caption{Comparisons of best-fit and $1\sigma$-range $f(z)$ (upper panels) and $\sigma_8(z)$ (lower panels) between CPDE and $\Lambda$CDM (left panels) and between CQDE and $\Lambda$CDM (right panels).}
        \label{fig:fonesigmas8onesigma}
    \end{figure}

    In figure~\ref{fig:fonesigmas8onesigma} we analyze the unmodified $f(z)$ and $\sigma_8(z)$ separately. 
    Faster growth rate means less dark matter in the past and explains the corresponding lower  amplitudes $\sigma_8$ for CPDE, which presents higher $f(z)$ compared to $\Lambda$CDM.
    The opposite happens in CQDE.
    The differences between the interacting models and $\Lambda$CDM appear to enhance as $z$ increases.
    The interacting models' $1\sigma$ ranges are consistent with $\Lambda$CDM except for CPDE's $1\sigma$-range $\sigma_8$, which is only marginally consistent with $\Lambda$CDM at low redshifts.

\begin{figure}[tb]
        \centering
        \includegraphics[width=\textwidth]{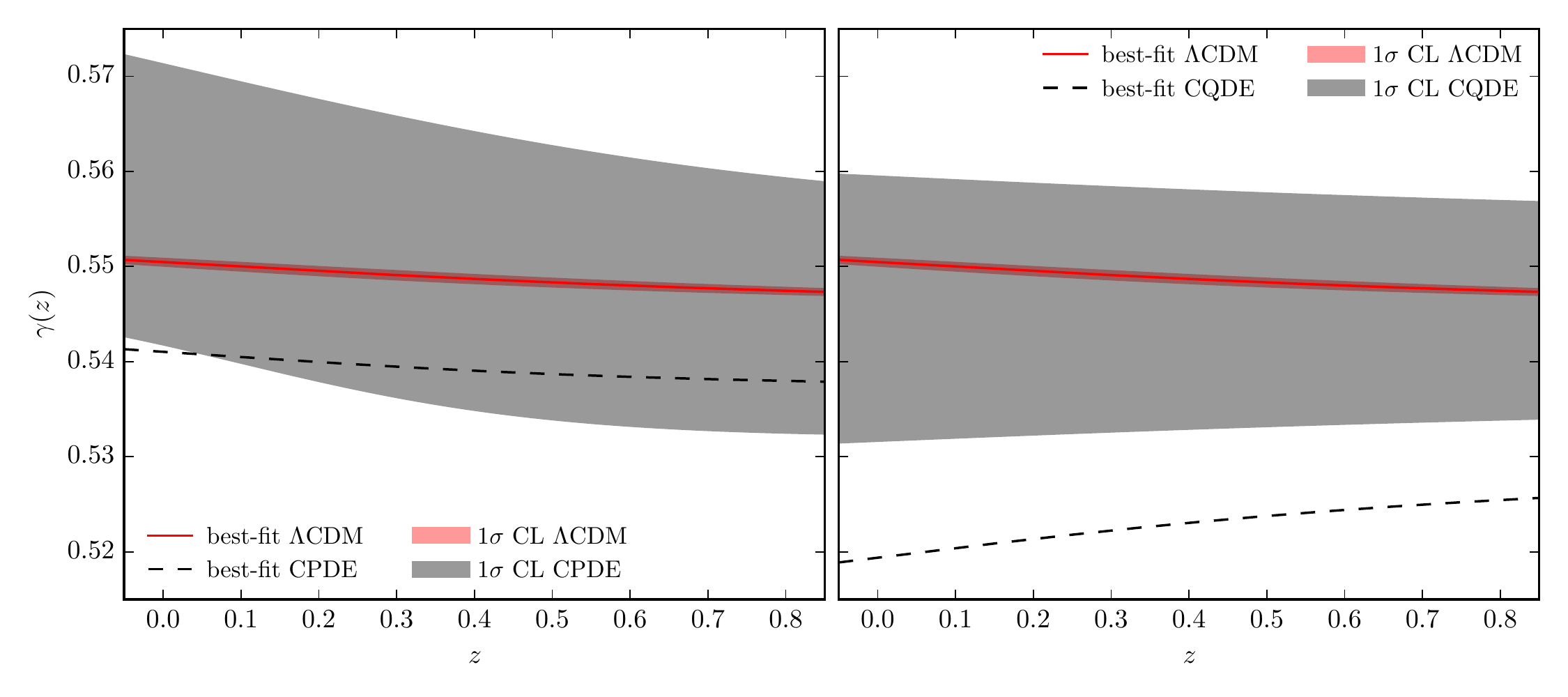}
        \caption{Comparisons of best-fit and $1\sigma$-range $\gamma(z)$ between CPDE and $\Lambda$CDM (left panel) and between CQDE and $\Lambda$CDM (right panel).}
        \label{fig:gammaonesigma}
    \end{figure}

    The $1\sigma$ range interval of $\gamma(z)$ in $\Lambda$CDM (see figure~\ref{fig:gammaonesigma}) is very tight because the only uncertainty involved is in the $\Omega_{\de,0}$ parameter, which is well constrained.
    The best-fit growth index is lower than $\Lambda$CDM's best-fit in the two \gls{cde} models, falling closer to $\Lambda$CDM in the CPDE case and outside its own $1\sigma$ range in the CQDE case. 
    However, their $1\sigma$ ranges are still consistent with $\Lambda$CDM in the redshift interval we are considering.

    \section{Conclusions}
    \label{sec:conclusions}
    In this work we have obtained the linear evolution equations for the perturbations of the dark matter fluid in cosmology with interaction in the dark sectors.
    We then rewrote those equations in terms of the growth rate, did the Taylor expansion around $\Omega_{\de} = 0$ and succeeded in deriving an analytic approximation for the growth index $\gamma$ when the interaction is proportional to $\rho_{\de}$. 
    The solution \eqref{eq:gammas} depends on the parameters of the dynamical \gls{de} equation of state and on the interaction coupling constant.
    The obtained analytic result for the growth index allows us to have the evolution of the growth of structures once the parameters of the coupled model are determined.
    Comparing with numerical calculations, we have proved that our analytic treatment works precisely within a region of the parameter space.
    The benefit of the analytic expression is that the influence of the coupling between dark sectors can be reflected clearly in the growth, which can help to distinguish from the models without interaction between \gls{dm} and \gls{de}.
    Furthermore, using the analytic growth formula, we can potentially reduce the computation time and confront our model to the observations more efficiently.
    In fact, in our analyses we were able to achieve convergence typically four orders of magnitude better, using a rather modest workstation, than in similar \gls{mcmc} analyses made with \texttt{CosmoMC} \cite{COSMOMC} and full numerical calculations in \texttt{CAMB}, running on a dedicated computer cluster, in comparable amounts of time.
    We have done the data fitting by confronting our model to the \gls{rsd} observations.

    On the other hand, we noticed that our analytic treatment is
    not generally effective. 
    For example, when the interaction between dark sectors is
    proportional to the energy density of \gls{dm}, the polynomial equation in $\Omega_{\de}$ forces the coupling to be always zero (see appendix~\ref{app:rhodmcase} for details).
    This problem can probably be solved if we can adopt a different parametrization of the growth rate.
    However, the generalization is not trivial and we leave it for further careful investigation in the future.

    Although theoretically we can predict the differences caused by the interaction between dark sectors in the growth of structures from the growth in other models, tight constraints on the model parameters, such as the \gls{de} \gls{eos} and the coupling, by using $f\sigma_8$ data exclusively are difficult to be obtained.
    This is mainly because of the low quality of the data at this moment.
    We expect that the situation can be improved soon in the next years with the advent of a new generation of powerful telescopes, e.g.~the SKA \cite{SKA2009}, BINGO \cite{Bingo2012,Bingo2014}, Euclid \cite{Euclid2013} and J-PAS \cite{JPAS2014} projects.

    \appendix
    \section{The case of an interaction proportional to the \glsentrytext{dm} density}
    \label{app:rhodmcase}
    We have also analyzed the case of an interaction proportional do the \gls{dm} density, ${Q_0}^{\dm} = - 3 \mathcal{H} \xi \rho_{\dm}$.
    We set the perturbed spatial part ${\delta Q_i}^{\dm} = 0$.
    The background evolution is given by $\dot {\bar\rho}_{\dm} + 3 \mathcal{H} \bar\rho_{\dm} = 3 \mathcal{H} \xi \bar \rho_{\dm}$.
    The perturbation equations \eqref{eq:dynamicalequations} with $\frac{\bar Q_0^{\phantom{0}\dm}}{\bar\rho_{\dm}} = \frac{\delta Q_0^{\phantom{0}\dm}}{\bar\rho_{\dm} \delta_{\dm}} = - 3 \mathcal{H} \xi$ are
    \begin{subequations}
        \begin{gather}
            \dot \delta_{\dm} + \theta_{\dm} = 0 \\
            \dot \theta_{\dm} + \left(1 + 3 \xi \right) \mathcal{H} \theta + \tfrac{3}{2} \mathcal{H}^2 \Omega_{\dm} \delta_{\dm} = 0
        \end{gather}
    \end{subequations}
    and give
    \begin{align}
        \label{eq:dmSOE}
        \ddot \delta_{\dm} + \left(1 + 3 \xi \right) \mathcal{H} \dot \delta_{\dm} - \tfrac{3}{2} \mathcal{H}^2 \Omega_{\dm}  \delta_{\dm}   = 0.
    \end{align}
    The functional form of this equation is even simpler than the \gls{cde} case of section~\ref{ss:CDEmodel} with respect to the standard evolution, with only one extra term proportional to $\xi$ in the coefficient of $\dot \delta_{\dm}$. 

    In terms of $f = \Omega_{\dm}^{\gamma}$, with $\frac{\ud \Omega_{\dm}}{\ud \ln a} = 3 \Omega_{\dm} \left[ \xi + w_{\de} \left(1 - \Omega_{\dm} \right) \right]$, the growth rate evolution equation is
    \begin{align}
        \label{eq:fgeneraleq}
        3 \left[ w_{\de} \left(1 - \Omega_{\dm} \right) + \xi \right] \frac{\Omega_{\dm}}{f} \frac{\ud f}{\ud \Omega_{\dm}} + f + \frac{1}{2} - \frac{3}{2} w_{\de} \left(1 - \Omega_{\dm} \right) + 3 \xi - \frac{3}{2} \frac{\Omega_{\dm}}{f} = 0,
    \end{align}
    which expanded in $\Omega_{\de}$ for $f = (\Omega_{\dm})^{\gamma_0 + \gamma_1 \Omega_{\de} + \ldots}$ gives the polynomial equation
    \begin{align}
        3 &\xi \left(1 + \gamma_{0} \right) 
        + \frac{1}{2} \left[3\left(1 - w_0 \right) - \left(5 - 6 w_0 \right) \gamma_0 + 12 \xi \gamma_1 \right] \Omega_{\de} + {} \nonumber \\
        \label{eq:M1poleq}
        {} &+ \frac{1}{4} \left[-{\gamma_0}^2 + 36 \xi {\gamma_2} +2 {\gamma_1} (12 {w_0}-5 - 3 \xi)+ \left(1 + 12 w_1 \right) \gamma_0 - 6 w_1 \right] \Omega_{\de}^2
        + \mathcal{O} ( \Omega_{\de}^3 ) = 0.
    \end{align}
    Unlike eq.~\eqref{eq:poleqCDE}, this now has a zero-th order part that does not vanish automatically and regardless of the interaction or other parameters as in the other model).
    In order for eq.~\eqref{eq:M1poleq} to hold, $3\xi \left(1 + \gamma_0 \right) = 0$ must be satisfied.
    This implies $\xi=0$, recovering the non-interacting results for $\gamma$ from the higher order terms, or $\gamma_0 = -1$ and $\gamma_1 = \frac{9 w_0 - 8}{12 \xi}$ (with $\xi \ne 0$) from the first-order coefficient, which does not seem to fit the observed growth unless perhaps with a fine tuning of the parameters.
    Also, note that this solution implies a non-smooth transition to zero interaction.
    Although numerically the growth rate in this model can still be, to some degree, well approximated by the power law $\Omega_{\dm}^{\gamma}$ form, as claimed in ref.~\cite{Linder2005}, analytically we can see that this form is not appropriate for a non-zero coupling in the interaction term that is proportional to $\rho_{\dm}$.

    \acknowledgments 
    This work is supported by CAPES, CNPq, FAPESP (Nos.~2011/18729-1, 2013/10242-1, 2013/26496-2), National Basic Research Program of China (973 Program 2013CB834900) and National Natural Science Foundation of China.

    \bibliographystyle{JHEP}
    \bibliography{refs.bib}

\end{document}

%% file: glossary.tex
\newacronym{de}{DE}{dark energy}
\newacronym{dm}{DM}{dark matter}
\newacronym{mcmc}{MCMC}{Markov Chain Monte Carlo}
\newacronym{lcdm}{$\Lambda$CDM}{$\Lambda$-Cold Dark Matter}
\newacronym{sneIa}{SNe~Ia}{Type Ia supernovae}

\newacronym{cmb}{CMB}{Cosmic Microwave Background}
\newacronym{bao}{BAO}{baryon acoustic oscillations}
\newacronym{wmap}{WMAP}{Wilkinson Microwave Anisotropy Probe}
\newacronym{ecmb}{eCMB}{extended Cosmic Microwave Background}
\newacronym{act}{ACT}{Atacama Cosmology Telescope}
\newacronym{spt}{SPT}{South Pole Telescope}
\newacronym{2df}{2dFGRS}{Two-Degree-Field Galaxy Redshift Survey}
\newacronym{rsd}{RSD}{redshift-space distortion}
\newacronym{flrw}{FLRW}{Friedmann-Lema\^itre-Robertson-Walker}
\newacronym{cl}{CL}{confidence level}
\newacronym{rms}{rms}{root-mean-square}
\newacronym{eos}{EoS}{equation of state}
\newacronym{efe}{EFE}{Einstein's field equation}
\newacronym{lss}{LSS}{large-scale structure}
\newacronym{cde}{CDE}{coupled dark energy}
\newacronym{fog}{FoG}{Fingers-of-God}
\newacronym{gr}{GR}{General Relativity}